\documentclass[prb,twocolumn,superscriptaddress,showpacs] {revtex4-1}
\usepackage{amsmath,amsfonts,amssymb,mathbbol,dsfont}
\usepackage{graphicx,psfrag,color}
\usepackage{dcolumn}
\usepackage{bm}
\usepackage{mathbbol, appendix}

\def\im{{\sf {i}}}
\def\x{{\bf x}}
\def\y{{\bf y}}
\def\r{{\bf r}}
\def\k{{\bf k}}
\def\d{{\bm \delta}}

\begin{document}

\preprint{}
\title{Equivalence of topological insulators and superconductors}
 
\begin{abstract}
Systems of free fermions are classified by symmetry, space dimensionality, 
and topological properties described by K-homology. Those systems belonging 
to different classes are inequivalent. In contrast, we show that by taking 
a many-body/Fock space viewpoint it becomes possible to establish equivalences 
of topological insulators and superconductors in terms of duality transformations. 
These mappings connect topologically inequivalent systems of fermions, jumping 
across entries in existent classification tables, because of the phenomenon of 
symmetry transmutation by which a symmetry and its dual partner have 
identical algebraic properties but very different physical interpretations.
To constrain our study to established classification tables, we 
define and characterize mathematically Gaussian dualities as dualities mapping
free fermions to free fermions (and interacting to interacting).  
By introducing a large, flexible class of Gaussian dualities we show that any 
insulator is dual to a superconductor, and that fermionic edge modes are dual 
to Majorana edge modes, that is, the Gaussian dualities of this paper preserve 
the bulk-boundary correspondence. Transmutation of relevant symmetries, particle 
number, translation, and time reversal  is also investigated in detail. As 
illustrative examples, we show the duality equivalence of the dimerized Peierls 
chain and the Majorana chain of Kitaev,  and a two-dimensional Kekul\'e-type 
topological insulator, including graphene as a special instance in coupling space, 
dual to a p-wave superconductor. Since Gaussian duality transformations 
are also valid for interacting systems we briefly discuss some such applications. 
\end{abstract}

\author{Emilio Cobanera}
\affiliation{Institute for Theoretical Physics, Utrecht University, 
3584 CE Utrecht, The Netherlands}
\email[Electronic address: ]{Coban003@uu.nl}
\author{Gerardo Ortiz}
\affiliation{Department of Physics, Indiana University, 
Bloomington, Indiana 47405, USA}

\maketitle

\section{Introduction}

In this paper we establish equivalences of topologically non-trivial
insulators\cite{review0,review1,review2} and superconductors. 
\cite{review2,alicea,beenakker} By means of duality transformations,
we show that any insulator has a dual superconducting partner, and 
the partners are either both topologically trivial or non-trivial.  We will 
focus on non-interacting dual partners, since general classification 
schemes exist for free-fermion systems.\cite{Altland, ten_fold_way,periodic_table,Stone2011} 
As it turns out, the duality transformations of this paper connect systems 
that are inequivalent from the point of view of these topological classifications. 
This is only possible because of the phenomenon of {\it symmetry transmutation}, 
by which a duality transformation maps a symmetry of one system with one physical 
interpretation, say particle number or time reversal, to a symmetry of the 
dual system with a different interpretation.
  
From an electromagnetic response viewpoint, insulating and 
superconducting phases of electron systems are  dramatically different. 
While the insulating phase is characterized by a vanishing current-carrying 
state at zero temperature, the superconducting phase supports a supercurrent 
and displays a perfect diamagnetic response, the Meissner effect. Yet there 
is a basic sense in which both states of matter are equivalent, since many of 
their defining properties stem from a common factor, that is, the existence of a 
gap in the bulk energy spectrum of  fermionic quasiparticles.  The additional 
presence of gapless, symmetry-protected, extended surface excitations 
defines operationally their topologically non-trivial character. One of the 
objectives of topological band theory is to classify, based on a few preferred 
(discrete) symmetries and space dimensionality,  topologically distinct 
non-interacting (single-particle) Hamiltonians and their concomitant gapless 
edge excitations. 

For systems without gauge symmetries, duality transformations are implemented 
by unitary mappings,\cite{conprl, ours1} 
and so they preserve symmetries; the symmetries of a system are in one-to-one
correspondence with the symmetries of its dual partner. However, the physical
interpretation of a symmetry and its dual image can be markedly different. 
Holographic symmetries\cite{holographic} constitute a most extreme example. 
For some pairs of dual partners, one of the systems displays boundary symmetries,
mapped to global symmetries of its dual partner. In this case we 
call the boundary symmetry holographic. This phenomenon is remarkable 
because a symmetry that is formally lost in the thermodynamic limit, the 
holographic symmetry, is mapped by duality onto a symmetry that may 
become spontaneously broken in that limit. Because of this, 
not uncommon, example of symmetry transmutation, it is conceivable 
that a duality  may map a particle conserving system to a 
non-conserving one, simply by mapping the \(U(1)\) symmetry of particle number 
to a dual \(U(1)\) symmetry that does not have that interpretation. 

While these arguments are encouraging in the search for equivalences of 
insulators and superconductors, there are at least two other obstacles besides
particle (non)conservation. First, in general, dualities for fermions will not 
preserve the quadratic (or Gaussian) character of a model system, often mapping 
free-fermion systems to interacting ones. Second, for topologically
non-trivial systems, even if one were to find dualities matching non-interacting 
dual partners, there is in general no reason to expect that these dualities should
also preserve the locality properties of the quasiparticle modes. For example, 
at zero energy, modes localized at a boundary may also be interpreted as 
boundary symmetries. Extrapolating from the experience with holographic 
symmetries, one would expect these modes to become delocalized after a 
duality transformation. 

As it turns out, both obstacles may be overcome, and, as a consequence, 
there is no fundamental obstruction to the construction of equivalences of 
topological insulators and superconductors in terms of dualities. 
Section  \ref{free_dualities} introduces the special class of duality maps 
that establishes those equivalences. The starting
point is the characterization of  duality transformations that preserve 
the quadratic fermionic nature of a given model system; we will 
call this transformation {\it Gaussian duality}. Next we will construct 
a large class of such dualities in any number of spatial dimensions, in order
to create a toolkit for generating a topological superconductor from any 
given topological insulator in a systematic fashion. In other words, given 
a topological insulator one can always find at least one dual topological 
superconductor associated to it. This process is, of course, 
reversible since the duality transformation is an isometry. Hence our 
results strongly suggest that there may exist equivalences
of topological insulators and superconductors across all entries of the 
topological classification table, at least for constant 
space dimension. Dimensional reduction by dualities is possible,
\cite{exact_dim_red} but we will not obtain any Gaussian instance of 
this phenomenon in this paper. Section  \ref{free_dualities} ends 
with the fundamental concept of symmetry transmutation as applied to 
fermion parity, translation and time-reversal symmetries. In particular, 
we will find a quantitative  connection between changes in translation 
symmetry and breaking of particle conservation. 

Particular and emblematic examples include the proof that the insulating 
dimerized Peierls \cite{Peierlsbook,SSH} and superconducting Kitaev (at 
vanishing chemical potential)\cite{kitaev_wire, Lieb1961} chains are dual partners, 
and the equivalence of graphene to a popular example \cite{diez2014, milsted2015} 
of a weak\cite{bernevigbook} topological superconductor in two spatial dimensions. 
It is in Section \ref{examples} that we present these two  prototypical equivalences.
These dual partners do not simply {\it resemble} each other, but  are 
isospectral from a many-body standpoint, for finite lattices and various 
boundary conditions. No doubt, this fact seems odd at first sight, since 
the spinfull Peierls chain for example partially breaks translation symmetry but 
not (the standard) time reversal or particle conservation, while its dual partner, Kitaev's 
Majorana chain, breaks time reversal and particle conservation but not 
translation symmetry. (In Appendix \ref{appA} we derive the superconducting 
dual of the \(m\)-merized Peierls chain, and discuss the differences 
between $m$ odd and even.) Similarly, graphene displays the 
symmetry of the honeycomb lattice\cite{Wallace_graphene} while it dual  superconducting 
partner sits on a square lattice. Remarkably, the Gaussian duality allows 
us to qualitatively understand the difference between zig-zag and 
armchair terminations in graphene.\cite{graphene_edges} The explanation to all these 
seemingly paradoxical observations is symmetry transmutation. 
In Appendix \ref{dualbcs} we describe a simple Gaussian duality
mapping a ($s$-wave) BCS superconductor to an insulator 
in any number of dimensions.

Another important issue addressed in Section \ref{examples} is the 
locality character of our Gaussian dualities, i.e., the  problem of 
showing that localized zero-energy modes are mapped to dual 
zero-modes that are also localized.   It is remarkable to have the 
possibility to generate {\it localization-preserving} Gaussian dualities.
In other words, there are no holographic symmetries associated to the 
Gaussian dualities of this paper: global symmetries map to global symmetries, 
and the localization properties of energy modes are also preserved, though 
edge modes may be shuffled among boundaries. In particular, 
the zig-zag boundary of graphene is exactly dual to the 
``Kitaev edge''  of Refs. \onlinecite{diez2014,milsted2015}. 
We show how topological defects and edge states map, 
and also how the nature of those excitations transmutes from 
(canonical) fermionic to Majorana character  by duality. 
Interestingly, we analytically construct {\it exact} (as opposed to 
asymptotic) zero-energy modes for any finite length 
Kitaev wire when the length is an odd number of lattice constants. 

An interesting outcome of our investigation is further confirmation 
that non-trivial topological quantum order is a property of a manifold 
of states interpreted relative to a given language
(a set of preferred observables\cite{ours3}); and not a property 
of the energy spectrum alone\cite{tqo} and some Hamiltonian 
singling out those states as energy eigenstates. We also investigate 
in Section \ref{examples} the interplay between dualities and topological 
invariants of the single-particle Hamiltonian. Indeed, our equivalences 
are duality mappings and hence necessarily isospectral.\cite{ours1}
However, some of these duality mappings  connect systems 
with ground  states characterized by different topological quantum 
numbers, thus belonging to different topological classes. For instance, 
Kitaev wire model belongs to the class D of the Dyson-Altland-Zirnbauer 
tenfold-way classification,\cite{Altland} 
while its dual, the dimerized (spinless) Peierls chain, belongs to the class AIII.
In the past, we have studied dualities mapping systems with 
topologically quantum-ordered ground-state manifolds to systems 
characterized by local (Landau) orders.\cite{ours1,holographic} 
 
In order to take advantage of duality transformations, it
is crucial to recognize that the many-body, and not the single-particle, 
representation of the system is the relevant one. 
There is absolutely no doubt that, from a computational standpoint, 
the single-particle representation (e.g., the Bogoliubov-de Gennes
equations) is the appropriate methodology to adopt in the non-interacting 
or mean-field case. Computationally, it reduces a problem of exponential 
complexity into one of polynomial complexity, thus allowing diagonalization 
of quite large system sizes. However, care must be exercised at the moment 
of analyzing properties like particle conservation that involve the whole 
many-body system. In particular, topological classification schemes and 
counting of many-body zero-modes relate directly to the many-body ground-state 
manifold. These cautionary remarks are entirely appropriate since Gaussian 
dualities connecting topologically non-trivial dual partners often seem at 
odds with one form or another of standard wisdom. However, they are 
entirely natural if one adopts the many-body (Fock-space) language of 
second quantization, and not the vector bundle analysis of single-particle 
Bogoliubov-de Gennes Hamiltonian matrices. 

Technically,  Majorana operators, defined (up to normalization) as the real and 
imaginary parts of the canonical fermionic field, generate a complex Clifford 
algebra naturally represented in Fock space, and our Gaussian dualities are 
characterized most naturally as isomorphisms of these Clifford algebras. The 
effect in single-particle (mode) space, where a different, exponentially smaller Clifford 
algebra emerges \cite{Altland,periodic_table} is induced {\it a posteriori}. Crucially, 
it follows that  our Gaussian dualities can also be used for investigating 
interacting many-body systems (see for example Section \ref{int_Peierls}).\cite{Ortiz2014}
Several different mathematical simplifications arise when Gaussian 
dualities are investigated in terms of Majorana operators. These simplify 
not only the search for equivalences, but also the analysis 
of symmetries, topological invariants and their transformation, and 
most importantly the mapping of boundary excitations. 

At this point it
becomes natural to ask about the extension of our work to bosons,
since it is clear that the notion of Gaussian duality applies to canonical
bosons just as well. However, the real and  imaginary parts of the bosonic
field satisfy the Heisenberg commutation relation, and so the 
theory of Gaussian dualities for bosons is bound to be markedly different 
from that of fermions. Due to this crucial technical difference, we defer 
the systematic study of bosonic Gaussian dualities to future research.
Nonetheless, we would still like to illustrate explicitly the point that 
symmetry transmutation is also operative in bosonic 
systems. Hence, in Appendix \ref{appB} we describe a duality mapping
of a bosonic Mott insulator to a quartet superconductor. We also comment 
briefly on the relevance of this example for cold atoms.

Section \ref{last} concludes with a summary and outlook. 

\section{Gaussian Dualities}
\label{free_dualities}

In statistical mechanics, duality mappings are unitary 
transformations that respect the locality structure of 
the many-body Hamiltonian or transfer matrix.\cite{conprl,ours1} 
In what follows we will set up the foundations to establish equivalences via 
dualities. Particularly, we will characterize operator maps relating 
Hermitian quadratic forms of fermions, i.e., Gaussian dualities, including the 
connection between such many-body 
dualities and the associated transformation of the single-particle 
Hamiltonian. Next we will introduce general techniques to decompose 
a very large class of free fermion models into sums of commuting 
Hamiltonians, and finally we will use this technique to construct 
a general class of Gaussian dualities.   

\subsection{What constitutes a Gaussian duality ?}

We are interested in establishing the conditions under which a duality 
transformation becomes Gaussian. 
Let us focus for simplicity on systems of free fermions defined on a 
lattice. Then, the most general free fermion Hamiltonian  is of the form 
\begin{eqnarray}
H=\sum_{i,j=1}^L\Big[K_{ij}c_i^\dagger c^{\;}_j
+\frac{1}{2}\Delta^{\;}_{ij}c_i^\dagger c_j^\dagger
+\frac{1}{2}\Delta^*_{ij}c^{\;}_jc^{\;}_i\Big],
\label{Hfreep}
\end{eqnarray}
with one-body and pairing interaction matrices
\begin{eqnarray}
K^\dagger=K,\quad \Delta^T=-\Delta ,
\end{eqnarray}  
where $^\dagger$ is the adjoint, $^T$  the transpose, 
 and $^*$ 
complex conjugation of a matrix. The creation (annihilation) operator of a fermion 
$c^\dagger_j$ ($c^{\;}_j$) in the single-particle orbital $\phi_j$, 
satisfying $\{c^{\;}_i,c^\dagger_j\}=\delta_{ij}$, is labelled 
by the generic subindex $j$ encoding arbitrary quantum numbers, 
including position, spin, orbital/band, angular momentum, etc. The total number of
single-particle orbitals is $L$. 

Equivalently, one can re-write $H$ in Nambu form
\begin{eqnarray}
H=\frac{1}{2} \, \alpha^\dagger \, h_{\sf BdG} \, \alpha +\frac{1}{2}{\sf Tr}\,K ,
\end{eqnarray}
where the column vector of fermion operators is given by
\begin{eqnarray} \hspace*{-0.5cm}
\alpha = \binom{c}{c^\dagger} , \mbox{ with } \alpha_j=c_j \ , \ \alpha_{L+j}=c^\dagger_j \ , \
j=1,\cdots, L ,
\end{eqnarray}
and the Bogoliubov-de Gennes single-particle Hamiltonian 
($2L \times 2 L$ matrix)  
\begin{eqnarray}\hspace*{-0.5cm}
h_{\sf BdG}&=&\begin{pmatrix}
K& \Delta\\
-\Delta^*& -K^*
\end{pmatrix} = \\
&& \hspace*{-1.0cm}
\im\mathds{1}\otimes  \Im(K)+\im\tau^x\otimes \Im(\Delta)
+\im \tau^y\otimes \Re(\Delta)+\tau^z\otimes \Re(K) \nonumber ,
\end{eqnarray}
where $\tau^\nu$, $\nu=x,y,z$, are Pauli matrices, and 
$\Re(\cdot) (\Im(\cdot))$ denotes the real (imaginary) part of the matrix. 
No matter what the specific matrices $K$ and $\Delta$ are, the 
single-particle Hamiltonian $h_{\sf BdG}$ always anticommutes with 
the antiunitary (particle-hole) operator 
\begin{eqnarray}
\mathcal{C}=\mathcal{K}\tau^x\otimes \mathds{1},\quad \mathcal{C}^2=\mathds{1} ,
\end{eqnarray}
i.e., $\{h_{\sf BdG},\mathcal{C}\}=0$, where \(\mathcal{K}\) denotes 
complex conjugation. That means that the single particle energy spectrum is 
antisymmetric with respect to its zero value, i.e., a particle-hole 
symmetric spectrum. By contrast, a chiral symmetry 
\(\mathcal{U}_{\sf chiral}\) is a unitary transformation that anticommutes
with \(h_{\sf BdG}\). For example, if \(\Im{(K)}=0=\Im{(\Delta)}\), 
then \(\mathcal{U}_{\sf chiral}=\tau^x\otimes \mathds{1}\). 

Suppose now that the unitary transformation \(\mathcal{U}_{\sf d}\) implements
a duality transformation,
\begin{eqnarray}
H^D=\mathcal{U}^{\;}_{\sf d} \, H \, \mathcal{U}_{\sf d}^\dagger ,
\end{eqnarray}
meaning that it transforms a local non-interacting Hamiltonian $H$ into another,  
dual $H^D$, that preserves the property of being also local. The map, 
however, could generate fermionic density-density interactions for instance. 
What are the general conditions under which $H^D$ is also an Hermitian quadratic form 
of fermions? 

To answer this question we will recast Hamiltonian $H$ 
of Eq. \eqref{Hfreep} in 
terms of Majorana operators 
\begin{eqnarray}
\gamma_{2j-1}=c^{\;}_j+c_j^\dagger,\quad \im \gamma_{2j}=c^{\;}_j-c_j^\dagger,
\end{eqnarray}
such that $\gamma^2_r=1$, $r=1,\cdots,2L$.
(The notation
\begin{eqnarray}
a_j=c^{\;}_j+c_j^\dagger,\quad \im b_j=c^{\;}_j-c_j^\dagger.
\end{eqnarray}
will be favored in later sections). Then
\begin{eqnarray}
H=\frac{\im}{2}\sum_{r,s=1}^{2L}h_{rs}\gamma_r\gamma_s+\frac{1}{2}{\sf Tr}\,K
\end{eqnarray}
becomes a quadratic form of Majorana fermions, with a  
 \(2L\times 2L\) matrix \(h\) that is real
and antisymmetric. 

We can now investigate the dual Hamiltonian. Remember that 
we want the duality map to be Gaussian, i.e., $H^D$ should also be
a quadratic form of Majorana fermions. A naive first, and trivial, attempt 
would be to keep the localization properties {\it identical}, i.e. 
\begin{eqnarray}
H^D=\frac{\im}{2}\sum_{r,s=1}^{2L} h_{rs}\gamma_r^D\gamma_s^D
+\frac{1}{2}{\sf Tr}\,K
\end{eqnarray}
where the dual operators are related to the originals as
\begin{eqnarray}
\gamma_r^D=\mathcal{U}^{\;}_{\sf d} \, \gamma_r \, \mathcal{U}_{\sf d}^\dagger,
\end{eqnarray}
and $\gamma_r^D$ is a Majorana fermion operator. 
This extreme local map, although Gaussian, is very restrictive and will not 
allow us to establish interesting equivalences between insulators and 
superconductors. We would like to relax the extreme locality constraint 
and allow for changes in the range of the dual matrix. In other words, 
we would like to realize a more general Gaussian duality
\begin{eqnarray}
H^D=\frac{\im}{2}\sum_{r,s=1}^{2L}h_{rs}^D \ \tilde{\gamma}_r \tilde{\gamma}_s+\frac{1}{2}{\sf Tr}\,K,
\end{eqnarray}
for some new Majorana operator \(\tilde{\gamma}_r\) significantly different
from the dual Majorana \(\gamma^D_r\) and the original one \(\gamma_r\). 

The argument above suggests setting up the relation
\begin{eqnarray}\label{linear_duality}
\mathcal{U}^{\;}_{\sf d} \, \gamma_r \ \mathcal{U}_{\sf d}^\dagger=\sum_{s=1}^{2L}
{O_{\sf d}}^s_r \ \tilde{\gamma}_s ,
\end{eqnarray}
so that the matrix \(O_{\sf d}\) may be computed explicitly as
\begin{eqnarray}\label{mb2sp}
	{O_{\sf d}}^s_r=
\frac{1}{2^L}{\sf tr}(\tilde{\gamma}_s \ \mathcal{U}^{\;}_{\sf d}\, 
\gamma_r \, \mathcal{U}_{\sf d}^\dagger).
\end{eqnarray}
Therefore, the duality map \(\mathcal{U}_{\sf d}\) is Gaussian if,
and only if, the matrix $O_{\sf d}$ is invertible, in which case it is also orthogonal. 
In the absence of a more educated choice, one may always set 
\(\tilde{\gamma}_s=\gamma_s\) in Eq. \eqref{linear_duality}.

The association \(\mathcal{U}_{\sf d}\mapsto O_{\sf d}\) shows that 
Gaussian dualities, though many-body in nature, also induce {\it a posteriori} 
a duality of the single-particle Hamiltonian 
\(h\) by the relation 
\begin{eqnarray}
h_{rs}^D=\sum_{r',s'=1}^{2L} h_{r's'} \ {O_{\sf d}}^r_{r'}{O_{\sf d}}^s_{s'}.
\end{eqnarray}
How much of the locality of the original system one preserves 
in the dual model will depend on the range of the matrix $O_{\sf d}$. 

Finally, let us contrast Gaussian dualities to other type of dualities 
that do not preserve the quadratic fermionic nature of the original theory. 
Consider the non-interacting Hamiltonian
\begin{eqnarray}
H=\im\sum_{j=1}^{L-1}[  t_x \, \gamma_{2j}\gamma_{2j+1}-t_y \, 
\gamma_{2j-1}\gamma_{2j+2}] ,
\end{eqnarray}
which  can be also described as a spin-1/2 Hamiltonian,
 \begin{eqnarray}
 H=-\sum_{j=1}^{L-1}[t_x \, \sigma^x_j\sigma^x_{j+1}+t_y\, \sigma^y_j\sigma^y_{j+1}] ,
 \end{eqnarray}
after the Jordan-Wigner map of Majorana operators
\begin{eqnarray}
\gamma_{2j-1}=\sigma^x_j\prod_{l=1}^{j-1}\sigma^z_l,\quad 
\gamma_{2j}=\sigma^y_j\prod_{l=1}^{j-1}\sigma^z_l ,
\end{eqnarray}
in terms of Pauli matrices $\sigma_j^\nu$, $\nu=x,y,z$.
A simple local rotation around the spin \(y\) axis,
\begin{eqnarray}
\sigma^x_j\mapsto \sigma^z_j,\quad \sigma^z_j\mapsto-\sigma^x_j \quad(j=1,\cdots,L),
\end{eqnarray}
induces a non-trivial change in the dual fermionic Hamiltonian. Although local, 
it is an interacting Hamiltonian
\begin{eqnarray} \hspace*{-0.6cm}
H^D=\sum_{j=1}^{L-1}[t_x\, \gamma_{2j-1}\gamma_{2j}\gamma_{2j+1}\gamma_{2j+2}
-\im t_y \, \gamma_{2j-1}\gamma_{2j+2}] .
\label{d_ham_1}
\end{eqnarray}
Therefore, the matrix $O_{\sf d}$ of Eq. \eqref{linear_duality} should fail to be invertible, 
and one can check that this is indeed the case. 

The dual Hamiltonian of Eq. \eqref{d_ham_1} has an interesting 
physical interpretation. It describes the competition 
between a $p$-wave superconducting chain in its topological phase 
and a density-density interaction. In the limit in which \(t_y\) vanishes, its ground 
state is number conserving, a Mott insulating state, otherwise 
its ground state is superconducting. For \(t_x=0\),
\(H^D\) has two exact zero-energy modes, \(\gamma_2\) and \(\gamma_{2L-1}\).
The evolution of these modes with \(t_x\) may be computed exactly
by exploiting the duality transformation connecting \(H^D\) to the free-fermion
Hamiltonian \(H\).

\subsection{Decoupling transformations}

The results of the previous section, especially the example at the
end of the section, show that generic dualities do not preserve 
the non-interacting character of a theory. Hence, to systematically 
establish equivalences of topological superconductors and insulators
it is necessary to determine all possible Gaussian dualities. Recall that the key difficulty
in searching for dualities is to identify unitary transformations that respect
the locality structure of the Hamiltonian, meaning that  Eq.\,\eqref{linear_duality} 
is not the full answer to our problem. We still need to address 
the issue of locality for the specific purpose of relating {\it topological 
insulators} to {\it topological superconductors}.

Let us denote by \(\r\) the sites  of a lattice $\Lambda$, i.e.,  \(\r\in \Lambda\) , 
defined in arbitrary space dimensions. 
A generic (second-quantized) electron system where the number of 
electrons $N$ is conserved is described by the Hamiltonian
\begin{eqnarray}
H=-\sum_{\r,\r',\sigma}\left [ t_{\r,\r'} \, c^\dagger_{\r,\sigma} c^{\;}_{\r',\sigma}+
{t}_{\r',\r} \, c_{\r',\sigma}^\dagger c^{\;}_{\r,\sigma}\right],
\end{eqnarray}
where $c_{\r,\sigma}^\dagger$ represents a canonical fermion 
creation operator at site $\r$ and spin $\sigma=\uparrow,\downarrow$. 
The hopping amplitude \(t_{\r,\r'}\) is related to \(t_{\r',\r}\) 
by complex conjugation, \(t^{\;}_{\r',\r}=t_{\r,\r'}^*\). 

Generically, the Hamiltonian above displays a broken time reversal
symmetry unless the hopping amplitudes are 
purely real or purely imaginary. For purely imaginary amplitudes, an 
internal decoupling occurs in the system that splits the particle 
conserving Hamiltonian \(H\) into four identical, independent and decoupled 
superconductors. The proof of this assertion relies on rewriting the 
particle conserving Hamiltonian $H$ above in terms of Majorana fermions
$a^{\;}_{\r,\sigma}$ and $b^{\;}_{\r,\sigma}$, such that
\begin{eqnarray}
a^{\;}_{\r,\sigma}=c^{\;}_{\r,\sigma}+c_{\r,\sigma}^\dagger,\quad 
\im b^{\;}_{\r,\sigma}=c^{\;}_{\r,\sigma}-c_{\r,\sigma}^\dagger,
\label{Majoranamap}
\end{eqnarray}
with the result
\begin{eqnarray}
H&=&-\sum_{\r,\r',\sigma}\left[ \left(\frac{t_{\r,\r'}-t_{\r',\r}}{4}\right)  
(a_{\r,\sigma} a_{\r',\sigma}+b_{\r,\sigma} b_{\r',\sigma}) \right . \nonumber \\ 
&&\left . + \,  \im \left (\frac{t_{\r,\r'}+t_{\r',\r}}{4}\right) 
(a_{\r,\sigma} b_{\r',\sigma}-b_{\r,\sigma} a_{\r',\sigma})
\right] .
\end{eqnarray}
Hence, if $t_{\r,\r'}$ is purely imaginary, 
$H=\sum_\sigma(\tilde{H}_{1,\sigma}+\tilde{H}_{2,\sigma})$ where the 
particle non-conserving Hamiltonians 
\begin{eqnarray}
\tilde{H}_{1,\sigma}&=&
-\frac{1}{2}\sum_{\r,\r'} t_{\r,\r'} \, a_{\r,\sigma} a_{\r',\sigma},\\
\tilde{H}_{2,\sigma}&=&
-\frac{1}{2}\sum_{\r,\r'}t_{\r,\r'} \, b_{\r,\sigma} b_{\r',\sigma},
\end{eqnarray}
commute, $[\tilde{H}_{1,\sigma},\tilde{H}_{2,\sigma'}]=0$. 

The four decoupled superconductors
\(\tilde{H}_{1,\uparrow},\tilde{H}_{2,\downarrow},\tilde{H}_{1,\uparrow},\tilde{H}_{2,\downarrow}\)
are isospectral: Any one of them can be mapped into any other one
by a local unitary transformation. On one hand, \(\tilde{H}_{1,\uparrow}\) 
(\(\tilde{H}_{2,\uparrow}\)) is mapped to \(\tilde{H}_{1,\downarrow}\)
(\(\tilde{H}_{2,\downarrow}\)) by a rotation in 
spin space. On the other hand, the unitary transformation 
\(\mathcal{U}_{\sigma}=\prod_\r \left(\frac{\mathds{1}+
a_{\r,\sigma} b_{\r,\sigma}}{\sqrt{2}}\right)\) 
maps \(\tilde{H}_{1,\sigma}\) to \(\tilde{H}_{2,\sigma}\).  
Hence it is possible to unequivocally associate \(H\) with a 
new Hamiltonian $H_{\sf reduced}$
\begin{eqnarray}\label{hreduced}
H\mapsto H_{\sf reduced}=
-\frac{1}{2}\sum_{\r,\r'} t_{\r,\r'} \gamma_\r\gamma_{\r'},
\end{eqnarray} 
of spinless Majorana fermions \(\gamma_\r\) that 
includes only one fourth of the original number of fermionic 
degrees of freedom, and represents any of the four Hamiltonians 
obtained by decoupling \(H\). 

The equivalence of Eq. \eqref{hreduced} states that the spectrum 
of \(H\) can be reconstructed from that of \(H_{\sf reduced}\). In 
particular, the zero-energy modes of the particle conserving Hamiltonian 
\(H\) are explained by the zero-energy modes of the spinless superconductor 
\(H_{\sf reduced}\). If \(H_{\sf reduced}\) has precisely \(s\) zero modes,
then \(H\) has precisely \(s^2\) zero modes. 

It is also possible to add spin interaction terms and retain 
some level of decoupling. For example, spin-orbit terms of the 
Rashba, Dimmock, or Dresselhaus type are linear in momentum and 
hence purely imaginary. Thus they couple \(\tilde{H}_{1,\uparrow}\) 
to \(\tilde{H}_{1,\downarrow}\) (and \(\tilde{H}_{2,\uparrow}\) to 
\(\tilde{H}_{2,\downarrow}\)), but they do not couple \(a\) to \(b\) 
Majoranas. In the presence of these types of spin terms, the decoupling 
transformation decomposes the particle conserving system into two 
rotationally-invariant superconductors with non-trivial spin dynamics.

Many of the Majorana lattice models investigated in the literature 
can be interpreted as the \(H_{\sf reduced}\) associated
to a particle-conserving Hamiltonian. The Majorana chain of Kitaev 
is obtained in one dimension from a chain with only nearest-neighbor 
hoppings.\cite{DasSarma2013} In two dimensions it is possible to obtain 
Kitaev's  Honeycomb model \cite{kitaev2006} (see Ref. \onlinecite{nussinov2009} for 
the immediate connection) as the reduced Hamiltonian associated
to the simplest\cite{castroneto09} model of graphene (see below), 
and the triangular Majorana lattice\cite{kraus2011} from a corresponding 
particle-conserving model on the triangular lattice. Decoupling on the 
square lattice obtains variations of Majorana arrays investigatigated in Ref. \onlinecite{diez2014}.   

An apparently less general but more often useful version of the decoupling 
transformation exits for purely real hopping amplitudes on a bipartite 
lattices \(\Lambda=A\cup B\), with generic lattice sites $\r \in \Lambda$. 
Let us denote by \(\x\in A\) and 
\(\y \in B\) the sites of each sublattice. The generic bipartite Hamiltonian 
\begin{eqnarray}
H=-\sum_{\x,\y,\sigma}
\left[ t_{\x,\y} \, c^\dagger_{\x,\sigma} c^{\;}_{\y,\sigma}+
{t}_{\y,\x} \, c_{\y,\sigma}^\dagger c^{\;}_{\x,\sigma}\right],
\label{Hhopp}
\end{eqnarray}
allows only for hopping from sublattice \(A\) to \(B\) or viceversa.
If $t_{\x,\y}$ is purely real-valued, that is, \(t_{\x,\y}=t_{\y,\x}\),
then
\begin{eqnarray}\label{bipartite_decoupling}
H=\sum_\sigma\left[{H}_{1,\sigma}-{H}_{2,\sigma}\right],
\end{eqnarray} 
where the superconducting Hamiltonians
\begin{eqnarray}
{H}_{1,\sigma}&=&
-\frac{\im}{2}\sum_{\x,\y} t_{\x,\y} \, a_{\x,\sigma} b_{\y,\sigma},\\
{H}_{2,\sigma}&=&
-\frac{\im}{2}\sum_{\x,\y} t_{\x,\y}\, b_{\x,\sigma}a_{\y,\sigma} , 
\end{eqnarray}
commute, $[{H}_{1,\sigma},{H}_{2,\sigma'}]=0$.
The spectral equivalence of  \({H}_{1,\sigma}\) and 
\({H}_{2,\sigma}\) is established by the unitary transformation
\begin{eqnarray}\label{for_chiral}
\mathcal{U}_{\sigma}=\prod_{\x,\y} 
\left (\frac{\mathds{1}+b_{\x,\sigma} a_{\x,\sigma}}{\sqrt{2}} \right)
\left (\frac{\mathds{1}+a_{\y,\sigma} b_{\y,\sigma}}{\sqrt{2}}\right )\, , 
\end{eqnarray}
that maps $H_{1,\sigma} \leftrightarrow H_{2,\sigma}$. 
An interesting corollary to Eqs. \eqref{bipartite_decoupling} and
\eqref{for_chiral} is that the unitary transformation 
\(\mathcal{C}=\mathcal{U}_{\uparrow}\mathcal{U}_{\downarrow}\) 
anticommutes with the Hamiltonian, \(\mathcal{C}H=-H\mathcal{C}\),
and so it defines a chiral symmetry in the sense that the spectrum of 
$H$ is particle-hole symmetric, i.e., for each positive eigenvalue 
$E_\alpha$ there exists a negative $-E_\alpha$. 

Just as before it is possible to associate a reduced Hamiltonian 
to \(H\). 

Consider, as an example, the chain of spinless fermions
\begin{eqnarray}\hspace*{-0.5cm}
H=-\sum_{j=1}^{2M}\left[t_j \, (c_j^\dagger c^{\;}_{j+1} +c_{j+1}^\dagger c_j^{\;})
+\, \epsilon_j(n_j-1/2)\right] , 
\label{generic_pc_chain}
\end{eqnarray} 
with quenched disorder in the real-valued hopping amplitudes \(t_j\) 
and on-site atomic energy \(\epsilon_j\). The chain 
has $L=2M$ lattice sites and periodic boundary conditions are 
assumed (\(c_{L+1}^\dagger=c_{1}^\dagger\)). Since the lattice is bipartite, 
\(H\) is the difference of two identical, independent superconductors, 
coupled by the on-site atomic energies. Let us associate a pair of 
Majorana fermions \(a_j,b_j\) to each site $j$, as in Eq. \eqref{Majoranamap}, 
and rewrite the Hamiltonian above in terms of Majorana degrees of freedom
\begin{eqnarray}
H&=&-\frac{\im}{2}\sum_{j=1}^{2M}
\left[t_j \, (a_{j+1}b_j-b_{j+1}a_j)+\epsilon_j\, a_jb_j\right]\nonumber\\
&&={H}_1-{H}_2+H_\epsilon,
\end{eqnarray}
with  commuting Hamiltonians
\begin{eqnarray}
{H}_1&=&-\frac{\im}{2}\sum_{j=1}^M \left[t_{2j-1} \,  a_{2j}b_{2j-1}-
t_{2j}\, b_{2j+1}a_{2j}\right] , \nonumber \\
{H}_2&=&-\frac{\im}{2}\sum_{j=1}^M \left[t_{2j-1} \, b_{2j}a_{2j-1}-
t_{2j} \, a_{2j+1}b_{2j}\right],
\end{eqnarray}
and
\begin{eqnarray}
H_\epsilon=-(\im/2)\sum_{j=1}^{M}[\epsilon_{2j-1} a_{2j-1}b_{2j-1}+
\epsilon_{2j} a_{2j}b_{2j}].
\end{eqnarray}

\subsection{A class of Gaussian duality transformations}
\label{general_ds}

We are now ready to introduce a
large class of Gaussian duality transformations.  Consider the cases 
for which the sublattices \(A\) and \(B\) of previous section are equivalent, 
meaning that there is a shortest, typically non-unique translation \(\d_1\) 
(\(\d_2\)) mapping sublattice \(A\) (\(B\)) to sublattice \(B\) (\(A\)). 
In set notation,
\begin{eqnarray}
\quad A+\d_{1}= B,\quad B+\d_2= A. 
\end{eqnarray} 
For the purpose of the duality transformation that we are about
to introduce it is often convenient to choose \(\d_1,\d_2\) to be
as parallel and short as possible. This condition guarantees
that the range of the hoppings in the dual Hamiltonian deviates
as little as possible from that of the original Hamiltonian.
A hypercubic lattice is simplest in that one may choose \(\d_1=\d_2\).

The mapping 
\begin{eqnarray}
a_{\y,\sigma}&\rightarrow& b_{\y+\d_2,\sigma},
\hspace*{1.05cm} \ a_{\x,\sigma} \rightarrow a_{\x,\sigma},  \nonumber \\
\qquad b_{\y,\sigma}&\rightarrow&b_{\y,\sigma}, \hspace*{0.7cm}  
\ \qquad b_{\x,\sigma} \rightarrow -{a}_{\x+\d_1,\sigma},
\label{duality_from_decoupling}
\end{eqnarray}
induces a unitary transformation that leaves \({H}_{1,\sigma}\) 
unchanged and transforms \({H}_{2,\sigma}\) as 
\begin{eqnarray}
{H}_{2,\sigma} \rightarrow 
{H}_{2,\sigma}^D=-\frac{\im}{2}\sum_{\x,\y} 
t_{\y-\d_1,\x-\d_2} \, b_{\x,\sigma}a_{\y,\sigma}. 
\end{eqnarray}   
In rearranging the sum over sites, we have assumed periodic 
boundary conditions or that the system is infinite. The dual 
superconducting Hamiltonian, 
\begin{eqnarray}
H^D&=&\sum_\sigma \left [{H}_{1,\sigma}-{H}_{2,\sigma}^D\right ]=\\
&-&\frac{\im}{2}\sum_{\x,\y,\sigma}\left[t_{\x,\y} \, a_{\x,\sigma}b_{\y,\sigma}-
t_{\y-\d_1,\x-\d_2} \, b_{\x,\sigma}a_{\y,\sigma}\right],\nonumber
\end{eqnarray}
can be rewritten in terms of creation and annihilation operators,
\begin{eqnarray}
H^D=-\sum_{\x,\y,\sigma}\Big[t^{\sf av}_{\x,\y}
(c_{\x,\sigma}^\dagger c^{\;}_{\y,\sigma}+c_{\y,\sigma}^\dagger c^{\;}_{\x,\sigma})
+\nonumber \\ \Delta_{\x,\y} 
(c_{\y,\sigma}^\dagger c_{\x,\sigma}^\dagger+c^{\;}_{\x,\sigma}c^{\;}_{\y,\sigma})\Big],
\label{Hdualgeneral}
\end{eqnarray}
where
\begin{eqnarray}
t^{\sf av}_{\x,\y}=\frac{t_{\x,\y}+t_{\y-\d_1,\x-\d_2}}{2}, \quad 
\Delta_{\x,\y}=\frac{t_{\x,\y}-t_{\y-\d_1,\x-\d_2}}{2}.\nonumber
\end{eqnarray}

Even though \(H+H_\epsilon\) is roughly as general 
as possible for a band electronic system of independent fermions, 
\(H^D+H_\epsilon^D\) remains a superconductor at vanishing chemical 
potential $\mu$. It is possible to include spin terms in \(H\) and
still obtain a dual superconductor featuring only local
interactions. Just as the duality breaks particle conservation
in general, we expect it to modify rotational properties since 
it has a highly non-trivial action on the operators of total spin. 
Since, and when, spin does not play any decisive role in the studied 
physical phenomenon we will drop it from the discussion in order to 
avoid confusing notation and obscure explanations.  

It is now straightforward to apply the general duality transformation 
of bipartite models to the disordered chain of the previous section. 
For this one-dimensional system, the mapping defined in 
Eqs. \eqref{duality_from_decoupling} reduces to 
\begin{eqnarray}
a_{2j-1}&\rightarrow& b_{2j},\hspace*{1.05cm} \ a_{2j} \rightarrow 
a_{2j}  \nonumber \\ b_{2j-1}&\rightarrow&b_{2j-1}, \hspace*{0.7cm}  
\ b_{2j} \rightarrow -{a}_{2j+1}  ,
\label{dualityPBC}
\end{eqnarray}
always identifying the index $L+1$ with $1$, and $0$ with $L-1$. 
Thus, while $H_1=H_1^D$ remains invariant, $H_2$ transforms as
\begin{eqnarray}
H^D_2=-\frac{\im}{2}\sum_{j=1}^{M} 
\left[t_{2j-2} \,  {b}_{2j}  a_{2j-1} -
t_{2j-1} \,   a_{2j+1}{b}_{2j} \right]
\end{eqnarray}
(the case $M=1$ is special in that the duality map keeps 
$H_2^D$ also invariant). The on-site atomic energy term transforms like
\begin{eqnarray}
&&H_\epsilon^D=
-\frac{\im}{2}\sum_{j=1}^{M}
\left[\epsilon_{2j-1}b_{2j}b_{2j-1}-\epsilon_{2j}a_{2j}a_{2j+1}\right]  \\
&&=\frac{\im}{2}\sum_{j=1}^{2M}
\left[\epsilon_j(c^\dagger_{j}c^{\;}_{j+1}-c_{j+1}^\dagger c^{\;}_{j})
+(-1)^j\epsilon_j(c^\dagger_{j}c^{\dagger}_{j+1}-c_{j+1}^{\;} c^{\;}_{j})\right ].
\nonumber
\end{eqnarray}
Combining all of these results, we obtain the dual superconductor
$H^D=H_1-H_2^D+H_\epsilon^D$
\begin{eqnarray}\label{dual_generic_chain}
H^D=
-\sum_{j=1}^{2M}\Big[\left (\frac{t_j+t_{j-1}-\im\epsilon_j}{2}\right )  c
^\dagger_{j}c^{\;}_{j+1}+\\
(-1)^j \left (\frac{t_{j-1}-t_j-\im\epsilon_j}{2}\right ) 
c^\dagger_{j}c^{\dagger}_{j+1}+{\rm H.c.}\Big ].\nonumber
\end{eqnarray}

\subsubsection{Symmetry transmutation: particle number and fermionic parity}
\label{transmutation1}

The duality transformation  Eqs. \eqref{duality_from_decoupling}, 
breaks particle conservation in general because the particle number 
(charge) operator
\begin{eqnarray}
\hat{N}=\sum_{\r,\sigma}\left[n_{\r,\sigma}-1/2\right]=\frac{\im}{2}
\sum_{\r,\sigma} a_{\r,\sigma} b_{\r,\sigma} .
\end{eqnarray}
associated to, and a symmetry of, \(H\) is drastically modified by 
the duality. Since 
\begin{eqnarray}
a_{\x,\sigma}b_{\x,\sigma}&\rightarrow& -a_{\x,\sigma}a_{\x+\d_1,\sigma}, \nonumber \\
a_{\y,\sigma}b_{\y,\sigma}&\rightarrow& -b_{\y,\sigma}b_{\y+\d_2,\sigma},
\label{incidentally}
\end{eqnarray}
the duality transformation maps \(\hat{N}\) to a symmetry 
\(\hat{N}^D\) of \(H^D\) that does not have the interpretation 
of a charge operator,
\begin{eqnarray}\hspace*{-0.6cm}
 \hat{N}^D=
-\frac{\im}{2}\sum_\sigma \left[ \sum_\x a_{\x,\sigma}a_{\x+\d_1,\sigma}+ \sum_\y
b_{\y,\sigma}b_{\y+\d_2,\sigma} \right],
\end{eqnarray}
while it is still true that \( [\hat{N}^D, H^D]=0\).

There is, however, a quantum number important from the point of view
of superconductivity that is {\it almost} preserved by duality:
fermionic parity. The operator of fermionic parity
\begin{eqnarray}
(-1)^F=e^{\im\pi \sum_{\r,\sigma} n_{\r,\sigma}}
= \prod_{\r,\sigma} (-\im a_{\r,\sigma} b_{\r,\sigma})
\end{eqnarray}
measures the parity of the total number of fermions. The BCS mean
field approximation breaks the symmetry of particle conservation
down to conservation of fermionic parity. The duality transformation 
maps
\begin{eqnarray}\label{fpdual}
&&(-1)^F\rightarrow\\
&& \prod_{\x,\sigma} (\im a_{\x,\sigma} a_{\x+\d_1,\sigma}) \prod_{\y,\sigma}
(\im b_{\y,\sigma} b_{\y+\d_2,\sigma})=(-1)^\sigma (-1)^F,\nonumber
\end{eqnarray}
where \((-1)^\sigma\) is the sign accumulated after permutations 
of the Majorana fermions to establish the original order \((-1)^F\).

Incidentally, Eq. \eqref{incidentally} shows that the duality 
\(H\rightarrow H^D\)  is more general 
in scope than the decoupling transformation that motivated it.
For example, adding an on-site energy term 
\begin{eqnarray}
H_\epsilon=-\sum_{\r,\sigma}\epsilon_\r (n_{\r,\sigma}-1/2)
\label{onsiteH}
\end{eqnarray}
(\(\r\in A\cup B\)) to \(H\), couples the reduced superconductors 
\({H}_{1,\sigma}\) and \({H}_{2,\sigma}\). It transforms as
\begin{eqnarray}\hspace*{-0.5cm}
H_\epsilon^D=\frac{\im}{2}\sum_{\x,\sigma}
\epsilon_\x a_{\x,\sigma} a_{\x+\d_1,\sigma}+
\frac{\im}{2}\sum_{\y,\sigma}\epsilon_\y  b_{\y,\sigma}b_{\y+\d_2,\sigma}.
\label{onsitedu}
\end{eqnarray}
Hence the effect of the on-site atomic energy term \(\epsilon_\r\) 
is to renormalize (by a purely imaginary amount) the hopping and 
pairing amplitudes of the dual superconductor in the directions 
$\d_1$, and $\d_2$.

\subsubsection{Translation Symmetry}

For the Gaussian dualities of Eq. \eqref{duality_from_decoupling}, the 
non-conservation of particle number for the dual partner \(H^D\) is 
precisely related to (partial) breaking of translation symmetry for \(H\), 
since the pairing potential \(\Delta_{\x,\y}\) vanishes if 
\(t_{\y-\d_1,\x-\d_2}=t_{\x,\y}\) (and \(H^D=H\) in this case). 
What is less obvious is that the translation symmetry of \(H^D\) may be
be higher than that of \(H\), in which case we are enlarging one group of 
symmetries (translations) at the expense of breaking another symmetry, 
particle conservation. This is explained by the transmutation of the translation
operation under duality.   

Let us focus for simplicity on a closed chain of spinless fermions. The extension
to more general settings is straightforward but notationally cumbersome. 
The map of Eq. \eqref{dualityPBC} leads to
\begin{eqnarray}
c_{2j-1}&\mapsto& c_{2j-1}^D=\frac{1}{2}(b_{2j}+\im b_{2j-1}),\nonumber \\
c_{2j}&\mapsto& c_{2j}^D=\frac{1}{2}(a_{2j}-\im a_{2j+1}), 
 \end{eqnarray} 
or, more explicitly, 
\begin{eqnarray}
c_{2j-1}^D&=&
\frac{1}{2}(c_{2j-1}-\im c_{2j})-\frac{1}{2}(c_{2j-1}^\dagger-\im c_{2j}^\dagger),
\nonumber  \\
c_{2j}^D&=&
\frac{1}{2}(c_{2j}-\im c_{2j+1})+\frac{1}{2}(c_{2j}^\dagger-\im c_{2j+1}^\dagger).
\label{dc1} 
 \end{eqnarray} 
These expressions show already transmutation of the translation operation, 
let us call it \(\hat T\). On one hand,
\begin{eqnarray}
{\hat T} c_j {\hat T}^\dagger =c_{j+1}\quad (j=j+L),
 \end{eqnarray} 
 and consequently (\({\hat T}^D=\mathcal{U}_{\sf d}^{\;} \, 
{\hat T} \, \mathcal{U}_{\sf d}^\dagger\)),
\begin{eqnarray}
{\hat T}^D c_j^D ({\hat T}^D)^\dagger =c_{j+1}^D\quad (j=j+L).
\end{eqnarray}
On the other hand,
\begin{eqnarray}
{\hat T} c_j^D{\hat T}^\dagger \neq c_{j+1}^D,
\end{eqnarray}
and so
\begin{eqnarray}
{\hat T}\neq {\hat T}^D.
\end{eqnarray}

This is the point to notice. If \({\hat T}\) happens to be a symmetry of 
\(H\), then \({\hat T}^D\) is necessarily a symmetry of \(H^D\). However, 
\({\hat T}^D\) cannot possibly have the interpretation of a translation by 
one site, since that physical interpretation continues to be attached to 
\({\hat T}\) ! (The action of \({\hat T}^D\) on the \(c_j\) may be computed by 
inverting Eqs. \eqref{dc1}.)  Notice, however,  by the 
same reasoning that
\begin{eqnarray}
{\hat T}^2=e^{\im \alpha_{\sf d}}({\hat T}^D)^2,
\end{eqnarray}
where the (possibly trivial) phase on the right-hand side is determined 
by the actual duality transformation. 

The concrete significance of this result will become apparent in the next section 
when we investigate the equivalence of the dimerized Peierls chain and the 
Majorana chain of Kitaev. What happens in that case actually is the following. 
The dimerized Peirls chain commutes with \({\hat T}^2\) and a very non-evident 
symmetry \(\mathcal{U}_{\sf d}^\dagger \, {\hat T} \, \mathcal{U}^{\;}_{\sf d}\) (not 
to be confused with 
\({\hat T}^D=\mathcal{U}^{\;}_{\sf d} \, {\hat T} \, \mathcal{U}_{\sf d}^\dagger\)). 
As a consequence, its dual partner (the Majorana chain) commutes with \({\hat T}\). 
This illustrates how Gaussian dualities may increase 
translation symmetry at the expense of breaking (transmuting) other
symmetries, e.g., particle conservation. 
 
It is revealing to to rewrite the Gaussian duality of Eq. \eqref{dualityPBC} 
as a map of fermions in crystal momentum space. The even-odd structure of the 
duality mapping evinced by Eqs. \eqref{dc1}  for example 
suggests that we should take a unit cell with two sites. To keep the 
notation simple, we will assume that \(L=2M\), with \(M\) odd. Then
\begin{eqnarray}\hspace*{-0.7cm}
c_{2j-1}=\!\!\!\sum_{l=-\frac{M-1}{2}}^{\frac{M-1}{2}} \frac{e^{-\im k j}}{\sqrt{M}} \, 
\hat{c}_{1,k},\quad 
c_{2j}=\!\!\!\sum_{l=-\frac{M-1}{2}}^{\frac{M-1}{2}} \frac{e^{-\im k j}}{\sqrt{M}} \, 
\hat{c}_{0,k} ,
\end{eqnarray} 
where \(k= 2\pi l/M\). Let us emphasize that we are not making any 
assumption about the symmetries of any particular Hamiltonian. We are just 
going to recast our duality transformation in a new light. With these definitions, 
the dual fermions in momentum space are 
\begin{eqnarray}
\hat{c}^D_{1,k}&=&\frac{1}{2}(\hat{c}_{1,k}-\im \hat{c}_{0,k})
-\frac{1}{2}(\hat{c}^\dagger_{1,-k}-\im \hat{c}^\dagger_{0,-k}),\\
\hat{c}^D_{0,k}&=&\frac{1}{2}(\hat{c}_{0,k}-\im e^{-\im k} \hat{c}_{1,k})
+\frac{1}{2}(\hat{c}^\dagger_{0,-k}-\im e^{-\im k} \hat{c}^\dagger_{1,-k}).\nonumber
\end{eqnarray}
The key point is that the dual fermions of momentum \(k\) are combinations
of the original fermions of momentum \(k\) and \(-k\). 
It follows that the induced single-particle duality \(O_{\sf d}\) is not block 
diagonal with respect to  momentum.

\subsubsection{Time-reversal Symmetry}

 The standard antiunitary operation \({\cal T}\) of motion reversal 
 may be specified by its action on the  creation and annihilation operators, 
\begin{eqnarray} \hspace*{-0.8cm}
{\cal T}\  c^{\;}_{j,\uparrow} \ {\cal T}^{-1}&=&c^{\;}_{j,\downarrow}, \hspace*{0.5cm}
{\cal T}\  c^{\dagger}_{j,\uparrow} \ {\cal T}^{-1}=c^{\dagger}_{j,\downarrow}, \nonumber \\ 
{\cal T}\  c^{\;}_{j,\downarrow} \ {\cal T}^{-1}&=&-c^{\;}_{j,\uparrow}, \ \  
{\cal T}\  c^{\dagger}_{j,\downarrow} \ {\cal T}^{-1}=-c^{\dagger}_{j,\uparrow},
\end{eqnarray}
with the important consequence that 
\begin{eqnarray}
{\cal T}^2=(-1)^{\hat{N}}.
\end{eqnarray}

The duality transformation Eq. \eqref{duality_from_decoupling} is spin
diagonal. Hence the dual fermions 
\begin{eqnarray}
c_{2j-1,\sigma}^D&=&
\frac{1}{2}(c_{2j-1,\sigma}-\im c_{2j,\sigma})-
\frac{1}{2}(c_{2j-1,\sigma}^\dagger-\im c_{2j,\sigma}^\dagger),
\nonumber  \\
c_{2j,\sigma}^D&=&
\frac{1}{2}(c_{2j,\sigma}-\im c_{2j+1,\sigma})+\frac{1}{2}(c_{2j,\sigma}^\dagger-\im c_{2j+1,\sigma}^\dagger),
 \end{eqnarray} 
are just as before, except for the additional  spin label. 
By construction, the dual antiunitary operation \({\cal T}^D=
\mathcal{U}_{\sf d}\, {\cal T} \, \mathcal{U}_{\sf d}^\dagger\), with
\begin{eqnarray}
({\cal T}^D)^2=(-1)^{\hat{N}^D},
\end{eqnarray}
acts as standard time-reversal on the dual fermions (for a discussion of 
\(\hat{N}^D\), see Section \ref{transmutation1}). One may check that 
\begin{eqnarray}
{\cal T}\neq {\cal T}^D,
\end{eqnarray}
and so there is transmutation of time-reversal symmetry. 
To put this result in perspective, suppose that both \(H\) and \(H^D\) commute
with \({\cal T}\). As we will see, this is the case for example in polyacetylene
and its dual superconducting partner (class DIII). Then, since \([H^D,{\cal T}^D]=0\), 
we have uncovered a (unitary) symmetry \({\cal T}^D{\cal T}\) 
of \(H^D\). 

An example of transmutation of time-reversal for canonical bosons 
(phonons) can be found in Ref. \onlinecite{ours1}, page 730.

\section{Equivalences of topological insulators and superconductors}
\label{examples}

Can Gaussian dualities in general, and in particular, the ones of this paper,
establish equivalences between topological insulators and topological 
superconductors? The discussion of the previous section shows
that there is no obstruction for this to be the case: many-body dualities can
jump across entries in single-particle classification schemes simply by 
transmuting key symmetries.  But symmetry transmutation is a necessary,
not a sufficient condition. For example, the duality of Eq. \eqref{dualityPBC}
maps the clearly trivial insulator,
\begin{eqnarray}
H=-\epsilon\sum_{j=1}^{2M}(n_j-1/2)
\end{eqnarray}
to the equally trivial superconductor,
\begin{eqnarray}
H^D&=&\frac{\im \epsilon}{2}\sum_{j=1}^{2M}
\Big[(c^\dagger_{j}c^{\;}_{j+1}-c_{j+1}^\dagger c^{\;}_{j})\\
&\ &\qquad\quad +(-1)^j(c^\dagger_{j}c^{\dagger}_{j+1}-c_{j+1}^{\;} c^{\;}_{j})\Big ],
\nonumber
\end{eqnarray}
in spite of the symmetry rearrangements it causes. 

For the Gaussian dualities of the previous section in particular, it is not 
hard to convince oneself that it must be the case that they map topologically 
(non)trivial systems to equally non(trivial) dual partners. In this section
we will study some paradigmatic examples of non-trivial partners. In one dimension,
we find that the dimerized Peirls chain and the Majorana chain of Kitaev
are dual partners, and we also investigate the mapping of 
topological defects under duality. The more general \(m-\)merized Peierls 
chain is investigated in the Appendix \ref{appA}.
In two dimensions, we study a topological insulator based on a 
Kekul\'e-like pattern of hopping matrix elements that includes graphene as a 
special case. It dual partner is a p-wave superconductor. In the limit where 
the insulator becomes graphene (a semi-metal), the dual superconductor reduces 
to a stack of Kitaev chains interconnected by pure kinetic hopping in the 
direction perpendicular to the chains. This superconducting realization of 
Dirac cones seems to be new in the literature. Appendix \ref{dualbcs} shows
an equivalence of a trivial BCS superconductor to a trivial insulator in any 
number of space dimensions.

\subsection{The Peierls chain is dual to the Kitaev chain}
\label{PeKi}

In one dimension, the dimerized Peierls chain at half-filling, 
proposed by Su, Schrieffer and Heeger (SSH) to model polyacetylene \cite{SSH},  
is the prototype of a topologically non-trivial band insulator, while 
the Kitaev chain is the prototype of a topologically non-trivial
superconductor. In spite of their physical differences, 
the Kitaev and Peierls chains are isospectral {\it in second quantization}, 
that is, as many-fermion systems. The reason is that there exists a 
Gaussian duality connecting both models. As we will see below, 
the mapping is different in 
detail for periodic and open boundary conditions. The Gaussian 
duality for open boundary conditions is crucial to understand the way 
boundary excitations are related, i.e.,  how the Majorana 
charge-neutral zero-energy edge modes of the Kitaev chain map into 
charged (canonical fermion) zero-modes of the Peierls chain. 

\begin{figure}[htb]
\includegraphics[width=1.0\columnwidth]{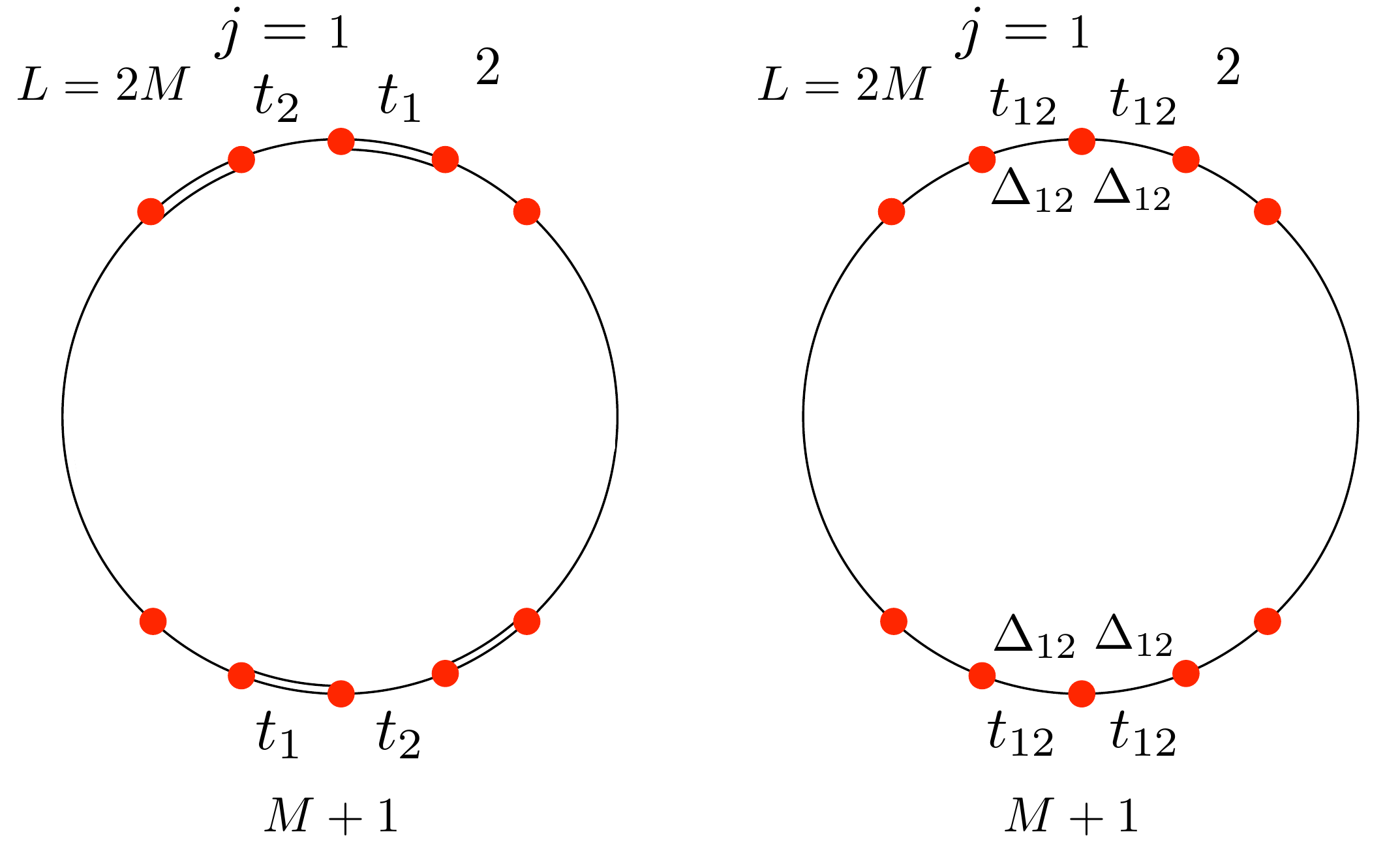}
 \caption{Peierls chain with periodic boundary conditions on the left and its dual 
 Kitaev chain superconductor on the right. Here, $L=2M$.}
 \label{Peierls-PBC}
\end{figure}

For simplicity,  and pedagogical reasons, in the following we will consider 
the spinless case. (The original model for polyacetylene\cite{SSH} involves 
spin-1/2 electrons and this fact is relevant for topological classification purposes
but it is not from the standpoint of Gaussian dualities that preserve spin, such 
as the ones defined in this paper.) 
Since the unit cell of the dimerized Peierls chain consists of two sites, 
the length (number of sites) \(L\) of the closed chain must be even, 
\(L=2M\). There are only two independent, periodically repeated hopping 
terms \(t_1\) and \(t_2\), see Fig. \ref{Peierls-PBC}. It follows
that the Peierls chain is a special case of the generic one-dimensional
Hamiltonian of Eq.\,\eqref{generic_pc_chain}, with 
\begin{eqnarray}
t_{2j-1}=t_1, \quad  t_{2j}=t_2,\quad \epsilon_j=0 , \  j=1,\cdots, L .
\end{eqnarray}
The dual of the Peirls chain,  
\begin{eqnarray}
\label{spinlessSC}
H^D&=&-\frac{\im}{2}\sum_{j=1}^{L} 
\Big[t_1 \, a_{j+1}{b}_j - t_2 \, {b}_{j+1}a_j\Big]  \\
&=&-\sum_{j=1}^{L} 
\Big[t_{12}  \, c^\dagger_{j}c^{\;}_{j+1} +  
\Delta_{12} \, c^\dagger_{j}c^{\dagger}_{j+1} + {\rm H.c.} \Big] \nonumber ,
\end{eqnarray}
is obtained by the same specialization of Eq.\,\eqref{dual_generic_chain},
with 
\begin{eqnarray}
t_{12}=\frac{t_1+t_2}{2},\quad \Delta_{12}=\frac{t_1-t_2}{2}.
\end{eqnarray} 
The Hamiltonian \(H^D\) is precisely the Majorana chain of Kitaev, 
at vanishing chemical potential $\mu$. 
The interplay of symmetries is noteworthy. While the dimerized 
Peierls chain shows reduced translation symmetry and conservation 
particle number, its dual the Kitaev chain has full translation 
symmetry and particle-number conservation is broken.
 
Next, we would like to comment on  polyacetylene (class AII), that is 
the case with real electrons (spinfull fermions). In this case, we obtain two 
copies of a Peierls chain, one for each spin component. 
Time-reversal symmetry is preserved and our Gaussian duality maps 
the SSH model to two copies of Kitaev's chain (class DIII), one for each spin component.

In Appendix \ref{appA} we analyze the superconducting equivalent of 
an $m$-merized Peierls chain with $m \geq 3$. 
There is a clear distinction between distortions with $m$ even and odd. 
In the latter case, the periodicity of the dual superconductor is 
doubled, and there are no zero-energy modes. 
For $m \geq 4$ even, the periodicities of the insulator and its 
dual superconductor are the same, in contrast to the dimerized case 
discussed above. It is well-known that for $m$ even there are zero-energy 
modes.

\subsubsection{Mapping of topological invariants}

A main goal of any topological classification of matter is to divide all 
possible quantum states of matter into equivalence classes. There is 
some degree of arbitrariness in the criteria used to define those 
classes. Once the criteria is established, e.g., by symmetry/dimension 
and reduced K-homology of bundles,\cite{periodic_table} two states in 
the same class are connected through a continuous map whose inverse is also 
continuous, i.e., a  homeomorphism.  To establish a characterization 
of the class one uses topological invariants, i.e., quantities that are 
preserved under the homeomorphism. 
That a particle-conserving system such as a Peierls insulator 
may be dual to a superconductor raises some conceptual 
issues for the topological classification of systems of free fermions.
What is the relation between the topological invariants characterizing 
these different but dual states of matter? The fermionic parity of 
the ground state of a superconductor such as Kitaev's, with $L=2M$ even, 
constitutes a good quantum number whose value depends on the boundary 
conditions. For periodic boundary conditions, the topologically  
non-trivial ground state of the Kitaev chain is non-degenerate and 
fermionic parity is odd, while it  is even
in the trivial phase.\cite{Ortiz2014} These facts hold independently 
of $M=L/2$ (the relevance of this remark will become clear below).
 
Fermionic parity may be computed in several equivalent ways. 
The topological character of this quantum number is revealed
by its connection to the Majorana number,\cite{kitaev_wire} 
\begin{eqnarray}
{\cal M}={\rm sgn}(\mu+2 t_{12}) \ {\rm sgn}(\mu-2 t_{12}) ,
\end{eqnarray}
defined as the product of signs of Pfaffians of an antisymmetric 
matrix at momenta $0$ and -$\pi$. This quantity identifies 
the topologically non-trivial phase as the one with ${\cal M}=-1$. 
On the other hand, we have seen that the dimerized Peierls 
chain maps into a Kitaev's chain at $\mu=0$. The Peierls chain
is always in a topologically non-trivial insulating phase, as long
as \(t_1\neq t_2\). However, the fermion parity of the Peierls' 
insulating non-degenerate many-body ground state is given by $(-1)^M$, 
and therefore it is defined by the parity of $M$, i.e., it can be odd 
or even depending on $M$. It is instructive to express the Peierls' 
chain in terms of Majorana fermions and compute the Majorana number, 
now with a doubled unit cell,  to realize that indeed ${\cal M}=(-1)^M$. 
The point is that fermion parity {\it is not} the good 
topological quantum number to characterize the Peierls' insulating 
phase despite the fact that it is exactly dual to Kitaev's chain model. 

From the point of view of the many-body duality transformation,
the mismatch is explained by the (very mild) transmutation of fermionic
parity, Eq.\,\eqref{fpdual}.

\subsubsection{Mapping of topological defects and boundary modes}
\label{peierls_kitaev_open}

\begin{figure}[htb]
\includegraphics[width=1.0\columnwidth]{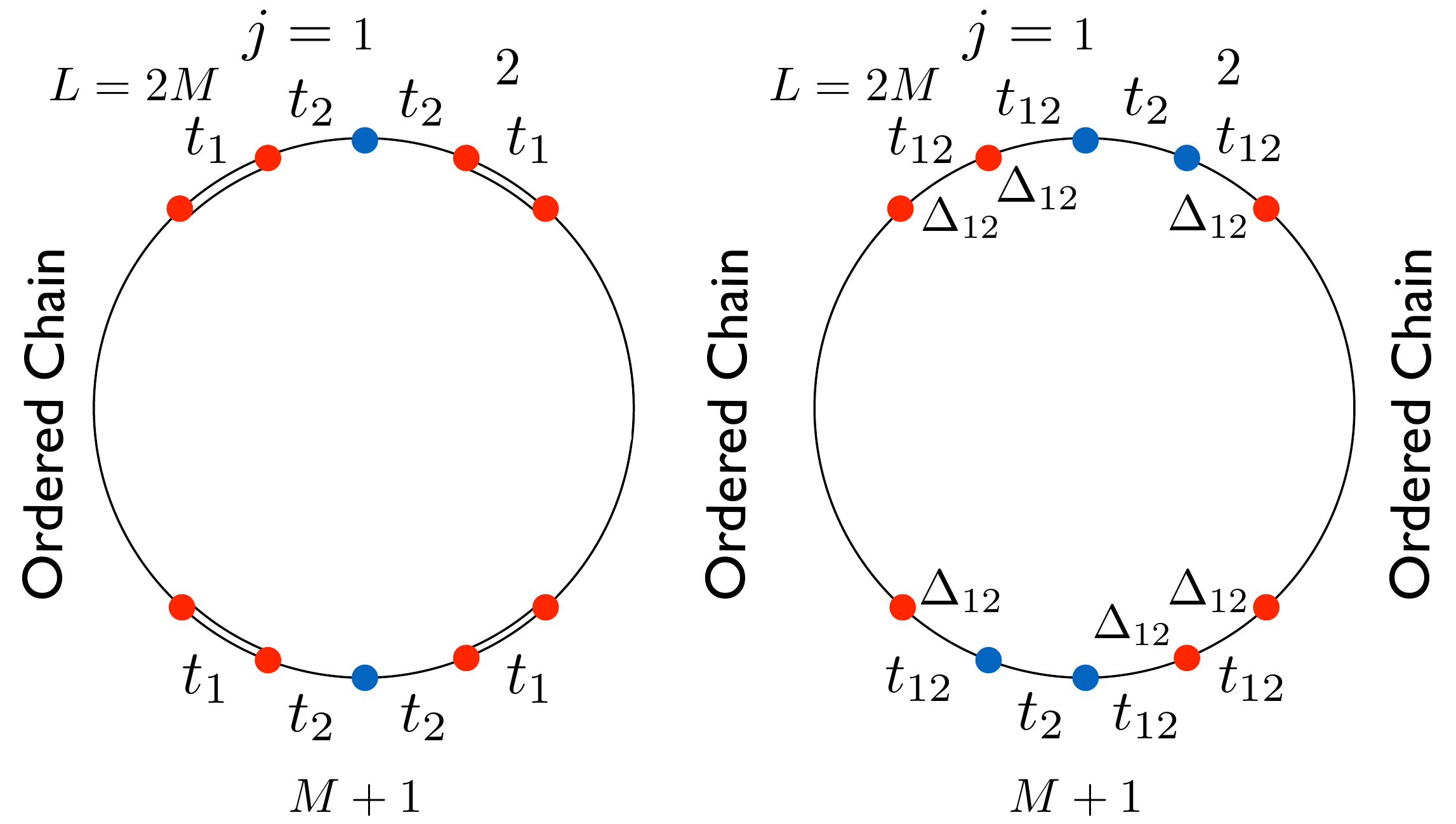}
\caption{Peierls chain (periodic boundary conditions) with a couple of 
defects on the left and its dual Josephson junctions of 
Kitaev chain superconductors on the right. Here, $L=2M$.}
 \label{Peierls-PBCDef}
\end{figure}

Consider for simplicity a dimerized, spinless, chain with periodic boundary 
conditions, $L=2M$ with $M \in$ odd, and two defects symmetrically 
 located at positions $j=1$ and $j=M$. This corresponds to the Hamiltonian
\begin{eqnarray}\label{HDefect}
H&=&- \sum_{j=2}^{M-1} t_{j-1({\sf mod}\, 2)}
(c_j^\dagger c^{\;}_{j+1} +c_{j+1}^\dagger c_j^{\;})\\
&&- \sum_{j=M+2}^{L-1} t_{j-(M+1)({\sf mod}\, 2)}
(c_j^\dagger c^{\;}_{j+1} +c_{j+1}^\dagger c_j^{\;}) \nonumber \\
&&-t_2(c_1^\dagger c^{\;}_{2} +c_{L}^\dagger c_1^{\;}+c_M^\dagger c^{\;}_{M+1} +
c_{M+1}^\dagger c^{\;}_{M+2} + {\rm H.c.}) ,\nonumber
\end{eqnarray}
where the first two terms represent two identical dimerized 
Peierls chains each of length $M-1$, and the last term represents the 
pair of defects, see Fig. \ref{Peierls-PBCDef}.  

Applied to the Hamiltonian \eqref{HDefect}, the duality 
transformation of the periodic chain, Eq.\,\eqref{dualityPBC},   
produces two dual superconductors coupled by two particle-conserving
segments,
\begin{eqnarray}
H^D&=&-\sum_{j=2}^{M-1} (t_{12}   c^\dagger_{j}c^{\;}_{j+1} +  
\Delta_{12}  c^\dagger_{j}c^{\dagger}_{j+1} + {\rm H.c.}  )\\
&&-\sum_{j=M+2}^{L-1} (t_{12}   c^\dagger_{j}c^{\;}_{j+1} -  
\Delta_{12}  c^\dagger_{j}c^{\dagger}_{j+1} + {\rm H.c.}  )
\nonumber \\
&&-t_2 (c_1^\dagger c^{\;}_{2} +
c_{M+1}^\dagger c^{\;}_{M+2} + {\rm H.c.}).
\end{eqnarray} 
Notice the change in sign of the superconducting order parameter 
across the links. Because of this phase difference, 
a Majorana zero mode is trapped at each weak link. 

For an infinite chain with a single defect located at the origin, 
one may use the same duality map, Eq. \eqref{dualityPBC} in order
to obtain a dual Majorana zero-mode localized at the origin. The 
defect that famously traps fractional charge  \(\pm e/2\), per spin direction,  in the Peierls 
chain is dual to a defect that traps a Majorana zero-mode!

\begin{figure}[htb]
\includegraphics[width=1.0\columnwidth]{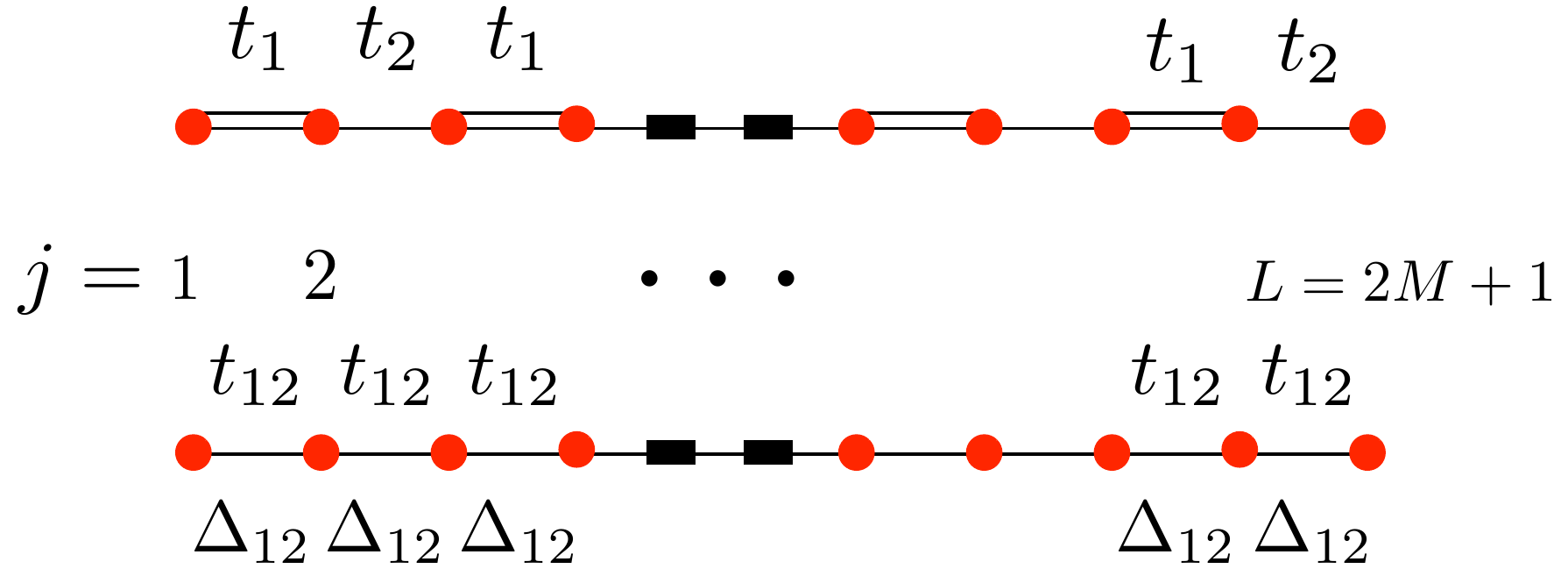}
 \caption{Peierls chain with open boundary conditions on the top and 
its dual Kitaev chain superconductor on the bottom. Here, $L=2M+1$.}
 \label{Peierls-OBC}
\end{figure}

For open boundary conditions, it is necessary to take the 
number of sites \(L=2M+1\) to be odd, see Fig. \ref{Peierls-OBC}. 
The duality transformation 
\begin{eqnarray}\label{duality_open}
a_{2j-1}&\rightarrow&a_{L-2(j-1)} \ , \ a_{2j} \rightarrow a_{2j}  
\nonumber \\
{b}_{2j-1}&\rightarrow&{b}_{2j-1} \hspace*{0.9cm} , 
\ {b}_{2j} \rightarrow {b}_{L+1-2j}  ,
\end{eqnarray}
leaves $H_1=H^D_1$ invariant, and transforms $H_2$ as follows
\begin{eqnarray}
H^D_2&=&-\frac{\im}{2}\sum_{j=1}^{M} (t_{2j} \,  
a_{L-2j}{b}_{L+1-2j}- t_{2j-1} \,  {b}_{L+1-2j} a_{L-2(j-1)}) ,
\nonumber \\
&=&-\frac{\im}{2}\sum_{j=1}^{M} (t_{L+1-2j} \,  
a_{2j-1}{b}_{2j}- t_{L-2j} \, {b}_{2j} a_{2j+1}) .
\end{eqnarray}

In the dimerized case there are only two different alternating 
hopping terms which satisfy
\begin{eqnarray}
t_{2j-1} = t_{L-2j}=t_1 \ , \ t_{2j} = t_{L+1-2j}=t_2 ,
\end{eqnarray}
with the end result that the dual total Hamiltonian represents a 
spinless superconductor
\begin{eqnarray}
H^D&=&-\frac{\im}{2}\sum_{j=1}^{L-1} (t_1 \, a_{j+1}{b}_j - t_2 \, {b}_{j+1}a_j)  \\
&=&-\sum_{j=1}^{L-1} \left (t_{12}  \, c^\dagger_{j}c^{\;}_{j+1} + \Delta_{12} \,  
c^\dagger_{j}c^{\dagger}_{j+1} + {\rm H.c.} \right )  \nonumber ,
\end{eqnarray}
that  is again the Kitaev chain Hamiltonian at vanishing, 
$\mu=0$, chemical potential. 

The Kitaev chain of this section (open boundary conditions, odd length,
and vanishing chemical potential) has two {\it exact} zero-energy  
modes, one per boundary point. The chain is reflection symmetric 
with respect to the central site $j=M+1$, and the many-body ground 
state is two-fold degenerate. The zero-energy mode associated to
the left boundary can be computed from the set of commutators 
\begin{eqnarray}
\ [-\im H^D, a_1]&=&t_2b_2, \nonumber \\
\ [-\im H^D, a_3]&=&t_1b_2+t_2b_4, \nonumber \\
\ &\vdots& \nonumber \\
\ [-\im H^D, a_{L-2}]&=& t_1b_{L-3}+t_2b_{L-1}, \nonumber  \\
\ [-\im H^D, a_L]&=& t_1 b_{L-1}. 
\end{eqnarray}
Let $\eta=t_2/t_1$, and assume without loss of generality $|\eta|<1$. 
From these commutators it is possible to show that the combination 
\begin{eqnarray}
\gamma^D_{\sf left}=
\frac{1}{\cal{N}}\sum_{j=0}^M\left (\! -\eta \right )^j \, a_{2j+1}
\end{eqnarray}
of \(a_j\) fermions at odd sites $j$ commutes with the Hamiltonian 
$H^D$. It constitutes an {\it exact symmetry} for any finite $M$.
The normalization factor is 
\begin{eqnarray}
{\cal N}(\eta)=\sqrt{\frac{1-\eta^{2(M+1)}}{1-\eta^2}}.
\end{eqnarray}

A similar calculation establishes the right symmetry
\begin{eqnarray} 
\gamma^D_{\sf right}=
\frac{1}{\cal{N}}\sum_{j=0}^M \left (\! -\eta \right )^j \, b_{L-2j}.
\end{eqnarray}
These exact Majorana zero-energy modes are exponentially localized.
If $|\eta|> 1$, the corresponding localized symmetries are obtained
by rescaling \(\gamma^D_\alpha \rightarrow (-1/\eta)^{M}\gamma^D_\alpha\),
and changing the normalization factor to  \({\cal N}(1/\eta)\). At 
\(|\eta|=1\) the mass gap vanishes in the thermodynamic limit $L \rightarrow \infty$. 

Notice the fundamental difference between chains of even or odd
lengths $L$. While it is possible to determine {\it exact} 
zero modes when $L$ is odd, this is not the case for $L$ 
even where the Majorana character of the edge modes is only {\it asymptotically 
exact} in the thermodynamic limit. The reason is simple. For {\it any} finite $L$, $H^D$ 
commutes with the global symmetries
\begin{eqnarray}
U_z=\prod_{j=1}^L(1-2n_j) \ , \ U_x =\prod_{j=1}^L \im (b^\dagger_j+b^{\;}_j) ,
\end{eqnarray}
where $b^\dagger_j=c^\dagger_j \prod_{l=1}^{j-1}(1-2n_l)$ is a hard-core 
boson. However, it is {\it only when 
$L$ is odd} that $\{U_z,U_x\}=0$. In turn, this implies that the 
whole many-body spectrum of $H^D$ is, at least, exactly two-fold degenerate.  
This symmetry analysis applies just as well to the more general Hamiltonian of 
Eq. \eqref{generic_pc_chain} with open boundary conditions, provided the on-site 
potential $\epsilon_j$ vanishes.

The duality transformation of Eq.\, \eqref{duality_open} maps
the boundary Majorana zero-modes of the Kitaev chain into corresponding 
boundary symmetries of the Peierls chain,
\begin{eqnarray} 
\gamma_{{\sf right},1}&=&\frac{1}{\cal{N}}\sum_{j=0}^M 
\left (\! -\eta \right )^j \, a_{L-2j}, \nonumber \\
\gamma_{{\sf right},2}&=&\frac{1}{\cal{N}}\sum_{j=0}^M 
\left (\! -\eta \right )^j \, b_{L-2j}. 
\end{eqnarray}
Unlike for the Kitaev chain, these two zero modes reside on one 
and the same edge. Hence, it is natural to recombine them into one 
exponentially localized fermionic mode,
\begin{eqnarray} 
c^\dagger_{\sf right}=\frac{\gamma_{{\sf right},1}- \im 
\gamma_{{\sf right},2} }{2} .
\end{eqnarray}
It is not surprising that both boundary symmetries of the Peierls chain
appear on one boundary point, and not both as in the 
Kitaev case. The reason is the lack of reflection symmetry about 
$j=M+1$ in the mapped Peierls chain. Nonetheless, its many-body ground 
state is two-fold degenerate, just as its dual Kitaev superconductor, 
indicating that the \(\mathds{Z}_2\) symmetry of fermionic parity
is odd in the thermodynamic limit also for the Peierls chain. This 
statement is of course confirmed by the exact solution of the Peierls 
Hamiltonian.


\subsubsection{Density-density interactions}
\label{int_Peierls}

As mentioned above Gaussian dualities can be used to establish 
equivalences in interacting many-body systems. Here, we illustrate 
this fact in another paradigmatic example. 
Consider the case of a dimerized Peierls chain (of length $L=2M$) 
at half-filling where electrons interact through a density-density 
Coulomb repulsion $V$
\begin{eqnarray}
H=\sum_{j=1}^{2M} 
\Big [&-&t_j \,(c_j^\dagger c^{\;}_{j+1} +c_{j+1}^\dagger c_j^{\;})
\nonumber \\
&+&\frac{V}{4}(1-2n_j)(1-2n_{j+1})\Big].
\end{eqnarray}
The duality transformation of Eq. \eqref{dualityPBC} maps the density 
operators as follows
\begin{eqnarray}
n_{2j-1}&\rightarrow&\frac{1}{2}(1- \im b_{2j-1} b_{2j}), \nonumber \\
n_{2j}&\rightarrow&\frac{1}{2}(1- \im a_{2j} a_{2j+1}) ,
\end{eqnarray}
and the resulting dual superconducting equivalent is given by
\begin{eqnarray} 
\label{spinlessSC}
&&\hspace*{-0.5cm}H^D
=-\sum_{j=1}^{L} \Big[t_{12}  \, c^\dagger_{j}c^{\;}_{j+1} +  
\Delta_{12} \, c^\dagger_{j}c^{\dagger}_{j+1} + {\rm H.c.} \Big] \nonumber \\ 
&&+ \frac{V}{4}\sum_{j=1}^{L} e^{\im \pi n_{j+1}} 
\Big[- c^\dagger_{j}c^{\;}_{j+2} +  
e^{\im \pi j} \, c^\dagger_{j}c^{\dagger}_{j+2} + {\rm H.c.} \Big]  ,
\end{eqnarray}
which clearly shows the competition and interplay between the band 
and Mott gaps.

A natural question that emerges is how robust is the 
Peierls phase to the presence of Coulomb interactions? The 
second, related, question is can interactions alone generate 
a topological Mott phase in the case where the non-interacting 
phase is metallic, i.e., $\Delta_{12}=0$? Since in the latter 
case  the model is exactly (Bethe ansatz) solvable, we know that 
for sufficiently large repulsion $V$, there exists a Mott phase. 

In Appendix \ref{appB} we analyze the phenomenon of 
symmetry transmutation in interacting boson systems. 

\subsection{Graphene is dual to a ``weak'' Topological Superconductor}

\begin{figure}[htb]
\includegraphics[width=.8\columnwidth]{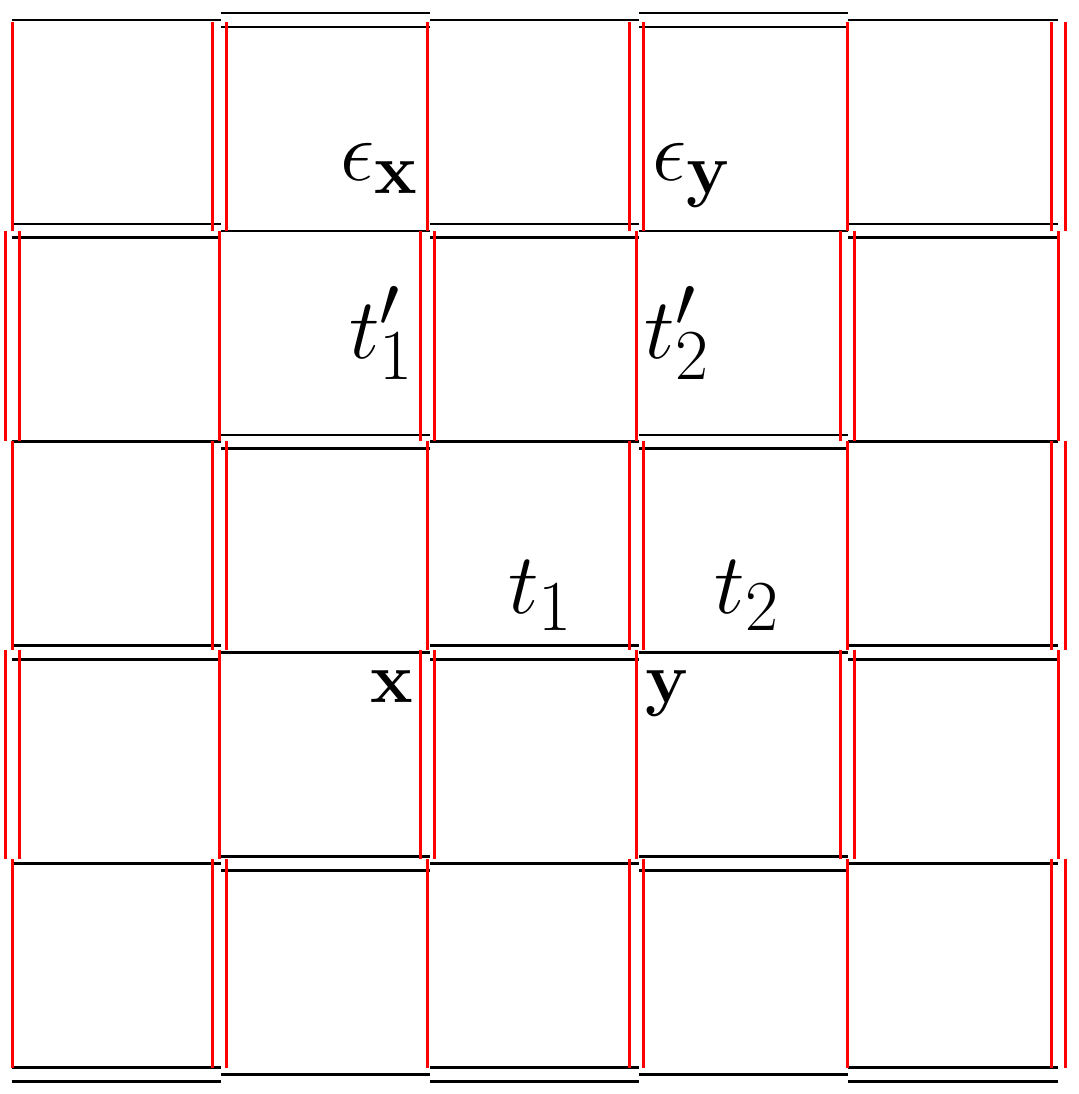}  

\centerline{${\Updownarrow}$}{\text{\large Dual Superconductor}}

\includegraphics[width=.8\columnwidth]{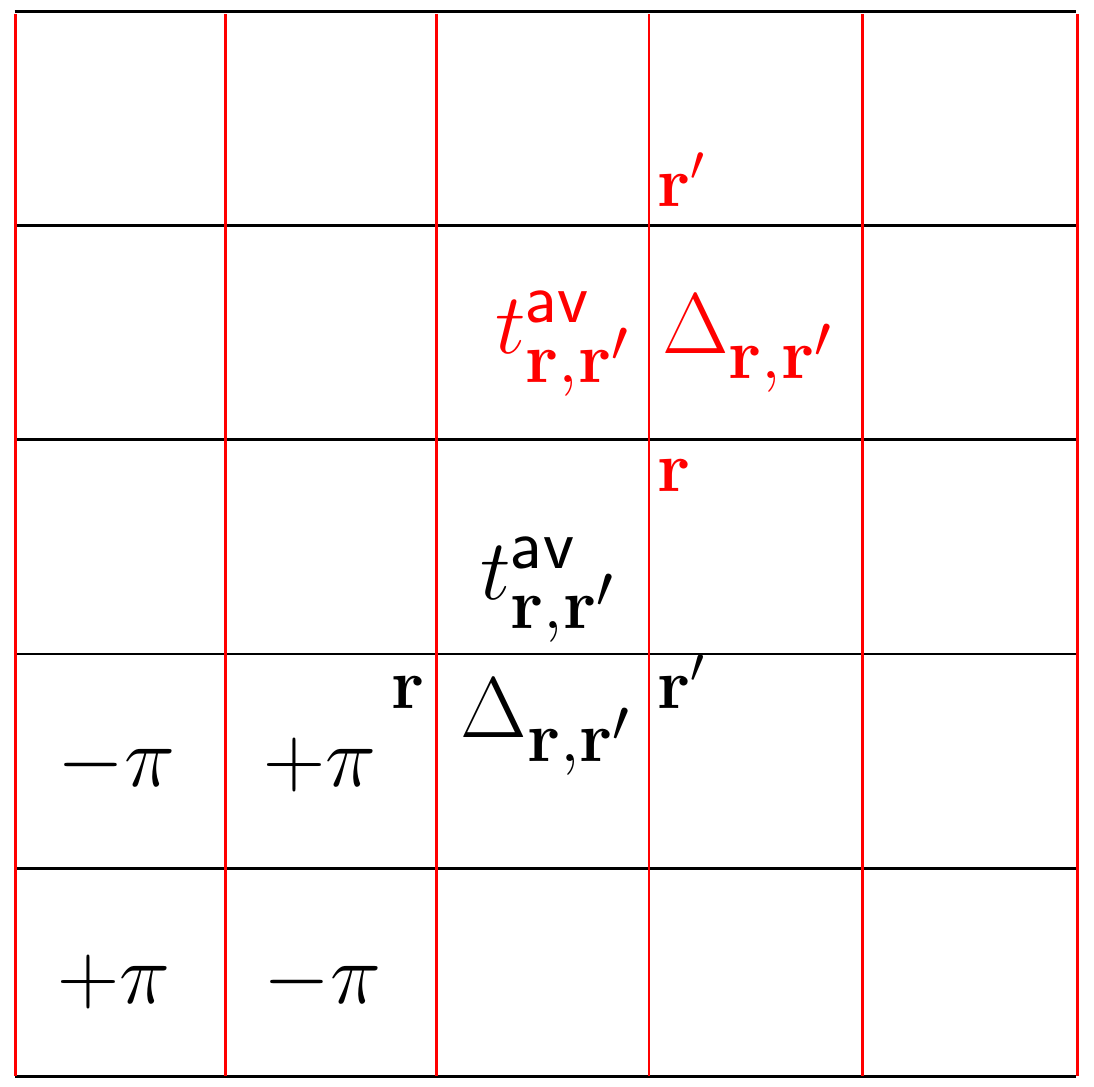}
 \caption{(Top) Topological insulator characterized by a particular 
Kekul\'e-type pattern of hopping matrix elements as depicted in the 
figure. Double bonds along the horizontal (vertical) direction represent 
the hopping matrix element $t_1$ ($t'_1$), while single bonds along 
the horizontal (vertical) direction represent $t_2$ ($t'_2$). The $\epsilon_{\x(\y)}$ 
are on-site atomic energies, constant on each sublattice. (Bottom) 
Dual chiral topological superconductor with $\pi$ fluxes per square 
plaquette distributed antiferromagnetically, and a $p_x+\im p_y$ 
superconducting order parameter. The dual relation between parameters 
is explained in the main text.}
\label{Pwave}
\end{figure}

In this section we establish a non-trivial equivalences in two space dimensions.
For conciseness, we investigate a  topological insulator on 
a square lattice, characterized by  a Kekul\'e-like pattern of hopping parameters 
\(t_1,t_1',t_2,t_2'\) and an on-site potential \(\epsilon_\x,\epsilon_\y\)
that is constant on each sublattice, see Figure \ref{Pwave}. In this way we 
manage to address several interesting models in a unified fashion.
The semimetal graphene for example is realized on the line  
$t_1=t'_1=t'_2$, $t_2=0=\epsilon_\x$. (A few other lines obtain graphene as well. Notice that the
honeycomb lattice is appears represented as a brick wall lattice in Figure \ref{Pwave}). 
As a consequence,\cite{Semenoff84} our model realizes a condensed matter 
analog of the (2+1)-dimensional parity anomaly\cite{Jackiw84}, and our 
duality transformation provides a superconducting dual representation of 
this phenomenon. As was the case in one dimension, the Gaussian duality 
mapping the insulating model to a topological p-wave superconductor
has the key property of preserving the locality of the edge mode excitations.

The two-dimensional underlying lattice $\Lambda$ considered in the following 
is bipartite with lattice points $\x=(x_1,x_2)$ and $\y=(y_1,y_2)$, such that 
$x_1+x_2 \in {\sf even}$ and $y_1+y_2 \in {\sf odd}$ integers. The total 
number of lattice points along the horizontal direction is $L_x$, and $L_y$ along 
the vertical direction, such that $L_x \times L_y$ defines the size of the lattice 
and where, for simplicity, $L_x$ and $L_y$ represent even integers.
Figure \ref{Pwave} (Top) is an example of a lattice $\Lambda$. 
Consider in particular the  lattice shown in Fig. \ref{Pwave} (Top) where, for a given 
point  $\x=(x_1,x_2)$, the corresponding hopping amplitudes of Hamiltonian 
 \eqref{Hhopp} are given by:
\begin{eqnarray}
t_{\x,\y}= \left \{
\begin{array}{cl}
t_1 & , \mbox{ for } \y=(x_1+1,x_2)  \cr
t_2 & , \mbox{ for } \y=(x_1-1,x_2) \cr  
t'_1 & , \mbox{ for }  \y=(x_1,x_2-1)  \cr
t'_2 & , \mbox{ for }  \y=(x_1,x_2+1)  
\end{array} \right .
\end{eqnarray}
and with on-site energies $\epsilon_\x=-\epsilon_\y$ (see Eq. \eqref{onsiteH}). 

This model, endowed with periodic (toroidal) boundary conditions, 
has a single-particle energy spectrum (bulk bands)  given by 
\begin{eqnarray}
E_{1,\k \sigma}&=&- \sqrt{\epsilon_\x^2 + A_{\k,+}^2+B_{\k,-}^2} \ , \ E_{4,\k \sigma}=- E_{1,\k \sigma} ,\nonumber \\
E_{2,\k \sigma}&=&- \sqrt{\epsilon_\x^2 + A_{\k,-}^2+B_{\k,+}^2} \ , \ E_{3,\k \sigma}= - E_{2,\k \sigma} , 
\end{eqnarray}
where the wavevectors $\k=(k_x,k_y)$  are defined in the Brillouin zone ($k_x=\frac{4 \pi}{L_x} n_x$, 
$k_y=\frac{4 \pi}{L_y} n_y$) with $n_x=0,1,\cdots,\frac{L_x}{2}-1$ and $n_y=0,1,\cdots,\frac{L_y}{2}-1$, and 
\begin{eqnarray}
A_{\k,\pm}&=&(t_1+t_2) \cos \left (\frac{k_x}{2} \right )\pm (t'_1+t'_2) \cos \left (\frac{k_y}{2} \right ) \nonumber \\
B_{\k,\pm}&=&(t_1-t_2) \sin \left (\frac{k_x}{2}\right ) \pm (t'_1-t'_2) \sin \left (\frac{k_y}{2}\right ) .
\end{eqnarray}
There is a chiral symmetry at work, since the energy levels are symmetrically 
distributed around zero energy and time reversal is not broken.  

The Gaussian duality of Eq. \eqref{duality_from_decoupling} with 
$\d_1=\d_2=(0,1)$ maps our Kekul\'e-type insulator into a $p$-wave 
superconductor with an antiferromagnetic distribution of $\pi$-fluxes 
per square plaquette (or more precisely, two spin copies of this system). 
The dual (chiral) superconductor is shown in 
Fig. \ref{Pwave} (Bottom) and corresponds to  (see Eqs. \eqref{Hdualgeneral} 
and \eqref{onsitedu}) 
\begin{eqnarray}\hspace*{-0.5cm}
H^D=-\sum_{\langle \r,\r'\rangle,\sigma}\Big[t^{\sf av}_{\r,\r'} \,
c_{\r,\sigma}^\dagger c^{\;}_{\r',\sigma}+(t^{\sf av}_{\r,\r'})^* \, c_{\r',\sigma}^\dagger c^{\;}_{\r,\sigma}
+\nonumber \\ \Delta_{\r,\r'} 
\, c_{\r',\sigma}^\dagger c_{\r,\sigma}^\dagger+(\Delta_{\r,\r'})^* \, c^{\;}_{\r,\sigma}c^{\;}_{\r',\sigma})\Big],
\end{eqnarray}
where $\langle \r,\r'\rangle$ represents nearest-neighbor links of a 
rectangular lattice with lattice points $\r=(r_1,r_2)$ and 
\begin{eqnarray}\hspace*{-0.5cm}
t^{\sf av}_{\r,\r'}= \left \{
\begin{array}{cl}
\frac{t_1+t_2}{2} & , \mbox{ for } \r'=(r_1+ 1,r_2)  \cr
\frac{t'_1+t'_2}{2}  - \frac{\im(-1)^{r_1+r_2}}{2} \epsilon_\r  & , \mbox{ for } \r'=(r_1,r_2+ 1) , 
\end{array} \right . \nonumber
\end{eqnarray}
\begin{eqnarray}
\Delta_{\r,\r'}= \left \{
\begin{array}{cl}
-\frac{t_1-t_2}{2} & , \mbox{ for } \r'=(r_1+ 1,r_2)  \cr
-\frac{t'_1-t'_2}{2} -\frac{\im}{2} \epsilon_\r  & , \mbox{ for } \r'=(r_1,r_2+1) 
\end{array} \right . .
\end{eqnarray}


\subsubsection{Mapping of topological boundary modes}

Our Kekul\'e-type insulator may displays zero energy modes if the on-site 
potential vanishes. It is instructive to consider explicitly the case of 
graphene to highlight the differences between zig-zag and armchair edge 
terminations\cite{ryu02} from the point of view of the dual superconductor. 
So let us take open boundary conditions along the 
$r_1$-direction and periodic along the $r_2$-direction, i.e., and open cylinder, and 
the parameter set $t_1=t'_1=t'_2$, $t_2=0=\epsilon_\x=\epsilon_\y$. 
Then, the Gaussian duality map used above for the toroidal boundary 
conditions (bulk) also works for the cylinder, since $\d_1=\d_2=(0,1)$ 
describes a translation along the periodic direction. One can see from
Fig. \ref{Pwave} that this situation corresponds to a zig-zag edge, while the 
parameter set $t_1=t_2=t'_1$, $t'_2=0=\epsilon_\x=\epsilon_\y$ would 
correspond to an armchair termination. The corresponding dual 
superconductors represent (two copies of) a stack of horizontal or vertical Kitaev 
chains respectively, in the topologically non-trivial regime. When the chains are
horizontal, they obtain the topological superconducting Majorana edge modes dual to
zig-zag terminated graphene. When the chains are vertical, the superconductor 
does not display edge modes, and its dual corresponds to the armchair terminated graphene. 

The general duality transformation of Section \ref{general_ds} breaks down for 
open boundary conditions in both $r_1$ and $r_2$ directions. Nonetheless, all 
of our conclusions hold just as well in this case were zig-zag and armchair terminations
coexist. In order to illustrate this point explicitly it becomes necessary to introduce 
a different Gaussian duality, showcasing once more the fact that exact dualities 
are very sensitive to boundary conditions. 

Let us focus for simplicity on a particular, spinless, case of our Kekul\'e-type
insulator with open boundary conditions in  both directions. The Hamiltonian 
of interest is given by
\begin{eqnarray}
H=H_{1}+H_{2}+H_v,
\end{eqnarray}
where
\begin{eqnarray}\hspace*{-0.5cm}
H_{1}&=&- \sum_{r_2=1}^{M_y}\sum_{r_1=1}^{M_x}\Big[t_1
\, c_{2r_1-1,2r_2-1}^\dagger c^{\;}_{2r_1,2r_2-1} \nonumber \\ 
&&\hspace*{1.6cm}+ t_2 \, c^\dagger_{2r_1,2r_2-1} 
c_{2r_1+1,2r_2-1}^{\;} + {\rm H.c.} \Big] ,
\end{eqnarray}
\begin{eqnarray}
H_{2}&=&-\sum_{r_2=1}^{M_y}\sum_{r_1=1}^{M_x}\Big[t_2 \, 
(c_{2r_1-1,2r_2}^\dagger c^{\;}_{2r_1,2r_2} \nonumber \\
&&\hspace*{1.6cm}+t_1 \, c^{\dagger}_{2r_1,2r_2} c_{2r_1+1,2r_2}^{\;}
+{\rm H.c.}\Big], 
\end{eqnarray}
$[H_1,H_2]=0$, and 
\begin{eqnarray}
H_{v}=-t_1\sum_{r_2=1}^{L_y-1}\sum_{r_1=1}^{L_x} \Big [  c^\dagger_{r_1,r_2}
c^{\;}_{r_1,r_2+1}+{\rm H.c.} \Big ] ,
\end{eqnarray}
with  $L_x=2M_x+1$ and $L_y=2M_y$, 
has an appealing interpretation as a stack of dimerized Peierls chains. 
Because the Peirls chain are alternating, the bulk translation symmetry 
in either direction is generated by translations by two sites. As a function 
of \(t_2\), the stack interpolates between a trivial metal, \(t_2=t_1\), 
and spinless graphene, \(t_2=0\), with zig-zag vertical and
armchair horizontal boundaries. 

The description of \(H\) as a stack of Peierls chains suggests a natural way
to map the system to a superconductor. Let us apply, to each horizontal chain, the 
duality of Section \ref{peierls_kitaev_open}, Eq. \eqref{duality_open}.
Unfortunately, this simplest alternative would obtain a non-local dual representation
of \(H_v\).  There is, however,  a way to fix this problem. Recall that the Gaussian
dualities of this paper are motivated by the observation that some systems may 
be split into independent subsystems, and then it is possible to rearrange one 
subsystem relative to the other. So, given the splitting of the open
Peierls chain into two subsystems, we could rearrange one subsystem for, 
say, the chains at odd height \(2r_2-1\), and the other subsystem for the chains 
at even height \(2r_2\). This idea is implemented by the Gaussian duality 
\begin{eqnarray}\label{duality_opennew}
a_{2r_1-1,2r_2-1}&\rightarrow&a_{L_x-2(r_1-1),2r_2-1} , \nonumber \\
{b}_{2r_1,2r_2-1} &\rightarrow& {b}_{L_x+1-2r_1,2r_2-1} ,\nonumber\\ 
a_{2r_1,2r_2} &\rightarrow& a_{L_x+1-2r_1,2r_2} , \nonumber  \\
{b}_{2r_1-1,2r_2}&\rightarrow&{b}_{L_x-2(r_1-1),2r_2} \label{duality_opennew2} .
\end{eqnarray}
The Majorana operators that are not explicitly listed remain unchanged.

Now, the alternating structure of the duality mapping in the vertical 
direction obtains the trivial transformation 
\begin{eqnarray}\label{constraint}
H_{v}^D=H_{v}.
\end{eqnarray}
The effect of the first half of the transformation on \(H_{1}\) follows
immediately from the work in Section \ref{peierls_kitaev_open}.
In \(H_2\), the hoppings \(t_1\) and \(t_2\) are exchanged relative to \(H_1\).
However, the second half of the transformation is also modified relative to the
first half in such a way as to precisely compensate for this exchange.  
Explicitly,
\begin{eqnarray}
H^D=H_1^D+H_2^D+H_{v},
\end{eqnarray}
with 
\begin{eqnarray}
H_1^D+H_2^D&=&-\sum_{r_2=1}^{L_y}\sum_{r_1=1}^{L_x-1}
\Big[t_{12}c_{r_1,r_2}^\dagger c_{r_1+1,r_2} \nonumber \\
&&\hspace*{0.8cm}+\Delta_{12}c_{r_1,r_2}^\dagger c_{r_1+1,r_2}^\dagger+{\rm H.c.}\Big].
\end{eqnarray}

As before, the dual system may be described as a stack of Kitaev wires,
but now with open boundary conditions in both directions. It realizes on 
its vertical boundaries the Kitaev edge,\cite{diez2014} a one-dimensional 
p-wave superconductor robust against statistical translation invariant 
disorder and/or interactions.\cite{milsted2015} We see that the zig-zag boundary of 
spinless graphene is dual to the Kitaev edge.

\section{Summary and outlook}
\label{last}

In this paper we have developed the general theory of Gaussian 
duality transformations for fermions, defined as maps that 
preserve a quadratic Hermitian form. The dual partner of a system of free fermions
is also a system of free fermions if the duality transformation 
is Gaussian, and both systems are equally local in space. The 
theory and practice of fermionic Gaussian dualities benefits 
from ``Majorana fermions," the complex Clifford algebra canonically
represented in Fock space. As a consequence, the theory of Gaussian 
dualities for canonical bosons is markedly different and left for 
a future publication.

As transformations of statistical mechanics, dualities are valued
for mapping strongly-coupled systems to weakly coupled ones. 
Gaussian dualities seem unconventional from this point of view, 
since systems of free fermions are, by definition, all weakly coupled.
To some extent, the conceptual mismatch is just a matter
of choice of language, that is, of physical representation of 
the systems under consideration.\cite{ours3} Take for example the simplest 
model of magnetic ordering, the transverse-field Ising chain. 
The self-duality  of this model, a 
non-local transformation of spins, is the prototype of a strong 
coupling/weak coupling duality transformation. However, the Jordan-Wigner 
mapping transforms the Ising model into the Majorana chain, and the 
self-duality of the Ising model into a Gaussian self-duality
of the Majorana chain.  

The key property shared by all duality transformations is
symmetry transmutation in face of the locality constraint.
This phenomenon occurs when a symmetry and its dual partner, 
necessarily a symmetry of the dual Hamiltonian, have different 
physical interpretations. For Gaussian dualities in particular 
we demonstrated transmutation of particle number, fermionic
parity, translation, spin rotation, and time-reversal symmetry
in various space dimensions. Transmutation of particle number is 
most conspicuous, since it allows for insulators (or semimetals) 
and superconductors (with zeroes of the gap function) to appear 
as dual partners.  

Because of symmetry transmutation, Gaussian dualities can
establish equivalences of topological insulators and superconductors,
relating systems classified as inequivalent from a single-particle
viewpoint. Here we  investigated
in detail two paradigmatic examples of such pairs of dual partners: 
the dimerized Peierls/Majorana chain (at vanishing chemical potential), and
graphene which happens to be dual to a weak topological 
superconductor. (Transmutation of lattice symmetry is manifest
in both examples. In particular, the superconductor dual to
graphene resides on a square lattice.) 

While our list of equivalences is far from exhaustive, our approach 
constitutes a general framework. Hence, future research should be
focused on searching systematically for other classes of Gaussian 
dualities besides the ones presented, aiming to obtain all possible 
equivalences across entries in established classification tables 
of electronic matter.
We can offer two comments as to what ``possible" should entail. First, 
we do not necessarilly expect equivalences of systems
of different space dimensionality, since dimensional reduction by
duality is possible but uncommon.\cite{exact_dim_red}  
Second, to obtain equivalences between topologically trivial and
non-trivial systems, it will become necessary to obtain the 
Gaussian analog of a holographic symmetry.\cite{holographic}
The Gaussian dualities of this paper do not realize holographic
symmetries: boundary symmetries (zero-energy modes) of a topologically 
non-trivial system are mapped to equally localized boundary symmetries 
of its dual partner, and topologically trivial systems are mapped to 
equally trivial dual partners. In short, our dualities preserve the 
bulk-boundary correspondence (for a useful discussion 
of this correspondence, see for example Ref. [\onlinecite{essin2011}] and 
references therein). 

Let us now discuss some possible applications of our results. 
It is interesting to think of dual partner systems, with at least
one open boundary, as scattering regions and ask what happens at the 
level of dualities if we attach leads to the system. The remarkable 
answer is that it is often possible to extend the Gaussian duality 
to apply to the whole system (scattering region plus leads), so that
its dual partner also has an unambigous interpretation as a scattering 
region with leads attached. Depending on the nature of the leads and 
Gaussian duality, the dual leads may be rearranged in space and/or 
become superconducting. In any case it is now possible to obtain 
quantitative dual mappings of transport properties to further 
extend the equivalence of insulators and superconductors. Notice
that it is no problem to add disorder, but symmetry transmutation
of the disorder ensemble should be expected. That is, if the disorder 
ensemble of one system displays some statistical symmetry, 
the dual system will be distributed according to a dual ensemble 
with a dual statistical symmetry of possibly very different nature.
These brief comments set the ground for investigating equivalences 
of statistical topological insulators and superconductors.\cite{fulga14}

Topologically non-trivial insulators and superconductors display 
zero-energy boundary modes associated to the degeneracy of their 
many-body ground-state energy level. In general, it is possible to 
trace the superconductor ground degeneracy back to some discrete 
symmetry left over from the breaking of particle conservation. 
This picture does not apply to insulators and so the origin of 
their topological degeneracy is harder to unveil. But our equivalence
can help here. In terms of its superconducting equivalent, the 
ground degeneracy of a topological insulator is explained by 
spontaneous symmetry breaking. Moreover, it can be detected and 
characterized numerically by investigating the system in terms
of Josephson-like physics. 
  
Finally, we would like to point out a very important
application of our duality approach to topological matter. 
Since our duality transformations are not  restricted to free 
fermion/boson systems (as shown in a couple of examples 
of this paper), it is apparent that they may become a powerful tool 
for a potential classification of interacting fermion/boson systems.

\acknowledgements

We gratefully acknowledge discussions with J. I. Cirac, M. Diez, G. van Miert, 
C. Morais Smith, C. Ortix, B. Seradjeh, and M. Tanhayi. This work is part of 
the DITP consortium, a program of the Netherlands Organisation for Scientific
Research (NWO) that is funded by the Dutch Ministry of Education, Culture and 
Science (OCW). GO thanks the Institute for Nuclear Theory at the University of 
Washington for its hospitality and the Department of Energy for partial 
support during the completion of this work.

\appendix

\section{ The $m$-merized Peierls chain}
\label{appA}

Let us describe briefly the superconducting model associated to the general 
$m$-merized Peierls case with $m>2$. Consider the $m=3$ (trimerized) Peierls 
chain with hoppings $t_1$, $t_2$, and $t_3$ and periodic boundary conditions 
($L=2M$ is divisible by $m=3$). Then, the dual superconducting Hamiltonian is 
given by
\begin{eqnarray}
H^D&=&-\sum_{j=1}^{L} \left ( w_j  \, c^\dagger_{j}c^{\;}_{j+1} + \Delta_j  \,
c^\dagger_{j}c^{\dagger}_{j+1} + {\rm H.c.} \right )   , 
\label{mrizeddual}
\end{eqnarray}
with coupling constants 
\begin{eqnarray}
w_1&=&\frac{t_1+t_3}{2}  \ , \  w_2=\frac{t_1+t_2}{2} \ , 
\  w_3=\frac{t_3+t_2}{2} \\
\Delta_{1,4}&=&\pm\frac{t_1-t_3}{2}  \ , \  
\Delta_{2,5}=\pm\frac{t_1-t_2}{2} \ , \  \Delta_{3,6}=
\pm\frac{t_3-t_2}{2}  \nonumber ,
\end{eqnarray}
and the rest of the couplings periodically repeat. Repeating 
the same line of reasoning, the case $m=4$ leads to a dual 
Hamiltonian such as Eq. \eqref{mrizeddual}, where the coupling 
constants are given by $w_1=\frac{t_1+t_4}{2},  w_2=\frac{t_1+t_2}{2} ,  
w_3=\frac{t_3+t_2}{2}  , w_4=\frac{t_3+t_4}{2}$, $
\Delta_{1}=\frac{t_1-t_4}{2}   ,  \Delta_{2}=\frac{t_1-t_2}{2}, \Delta_{3}=
\frac{t_3-t_2}{2}  ,   \Delta_{4}=\frac{t_3-t_4}{2}$. 
This clearly shows the difference between the cases where 
$m$ is odd from those where $m$ is even. In the latter case, 
the periodicity is always $m$ lattice constants, except for the particular 
dimerized $m=2$ case where the periodicity is one lattice 
constant, while the periodicity becomes $2m$ when $m$ is odd.   

\section{Insulating dual partner of the BCS superconductor}
\label{dualbcs}

There is an elementary Gaussian duality of an $s$-wave
BCS superconductor to a trivial insulator. Let us denote by \(\hat{c}_{\k,\sigma}\)
the annihilation fermionic field in momentum space $\k$, and assume 
$\epsilon_{\k}=\epsilon_{-\k}$. Then we may define the 
Gaussian duality 
\begin{eqnarray}
\hat{c}^{\;}_{\k,\uparrow}\mapsto \hat{c}^{\;}_{\k,\uparrow},\quad 
\hat{c}^{\;}_{\k,\downarrow}\mapsto \hat{c}_{-\k,\downarrow}^\dagger,
\end{eqnarray}
acting non-trivially only on the spin-down fermions. 
The mean-field BCS Hamiltonian  is
\begin{eqnarray}
\hspace*{-0.8cm}
H=\sum_{\k,\sigma}(\epsilon_{\k}-\mu) \,
\hat{c}_{\k,\sigma}^\dagger\hat{c}^{\;}_{\k,\sigma}-\sum_{\k}
\Delta \, (\hat c^\dagger_{\k,\uparrow} \hat c^\dagger_{-\k,\downarrow}+{\rm H.c.}) ,
\end{eqnarray}
and gets mapped to 
\begin{eqnarray} \hspace*{-0.5cm}
H^D=\sum_{\k} \big[
\epsilon_\k 
\hat{c}^\dagger_{\k,\alpha}\sigma_{\alpha,\beta}^z \hat{c}^{\;}_{\k,\beta}-h_\nu 
\hat c^\dagger_{\k,\alpha} \sigma_{\alpha,\beta}^\nu \hat c^{\;}_{\k,\beta}
\big]+C ,
\end{eqnarray}
with \(C=\sum_{\k}(\epsilon_\k-\mu)\), $\nu=x,y,z$, and  
\begin{eqnarray}
h_x=\Delta,\quad h_y=0,\quad h_z=\mu.
\end{eqnarray}
Particle conservation is restored in \(H^D\) at the expense of broken
(by symmetry transmutation) time-reversal and spin-rotation symmetry.

\section{A Bosonic insulator and its dual superfluid}
\label{appB}

In recent years, lattice models of bosons have acquired new 
relevance thanks to the spectacular experimental development 
of ultracold atom physics. In this section we will focus on a case of 
practical importance in which only three states
$\{|\bar{n}-1\rangle, |\bar{n}\rangle, |\bar{n}+1\rangle\}$ 
per lattice site $j$ are physically active. Then the particle 
operators \(g^{\;}_j,g_j^\dagger,n^{\;}_j\) act on these states as 
\begin{eqnarray}
&&g^{\;}_{j}|\bar{n}-1\rangle = 0 \ , \hspace*{1.2cm}  g^{\dagger}_{j}
|\bar{n}-1\rangle  = \sqrt{\bar{n}} \ |\bar{n}\rangle , \nonumber \\
&&g^{\;}_{j}|\bar{n}\rangle = \sqrt{\bar{n}} \ |\bar{n}-1\rangle  ,
\hspace*{0.45cm} g^{\dagger}_{j}|\bar{n}\rangle  = \sqrt{\bar{n}} \
|\bar{n}+1\rangle  , \nonumber  \\
&&g^{\;}_{j}|\bar{n}+1\rangle = \sqrt{\bar{n}} \ |\bar{n}\rangle  ,
\hspace*{0.45cm}  g^{\dagger}_{j}|\bar{n}+1\rangle = 0 ,
\end{eqnarray}
and
\begin{eqnarray}
&& n^{\;}_{j}|\bar{n}-1\rangle =  (\bar{n}-1)|\bar{n}-1\rangle, \nonumber \\
&& n^{\;}_{j}|\bar{n}\rangle = \bar{n}|\bar{n}\rangle, \nonumber \\
&& n^{\;}_{j}|\bar{n}+1\rangle =  (\bar{n}+1)|\bar{n}+1\rangle.
\end{eqnarray}
These are bosonic operators 
restricted to a finite(three)-dimensional Hilbert space per site. 
Their algebra and application to optical lattices\cite{optical_lattices}
 has been extensively investigated in 
Refs.\,\onlinecite{HMFT} and \onlinecite{ours3}. Here it will 
suffice to notice that there is a connection to spin \(S=1\) 
operators that, in the present paper and because of the duality 
we will apply later on,  we take to be
\begin{equation}\label{btos}
\sqrt{\frac{2}{\bar{n}}} \, g^\dagger_j =S^z_j+\im S^x_j=S^+_j , \quad 
n_j=S^y_j+\bar{n}. 
\end{equation}  

Here  we will focus on a system of bosonic atoms 
distributed with (integer) average particle density  $\bar{n}\geq1$  
on a one-dimensional optical lattice. We assume that the
dynamics of the system is described by the Hamiltonian 
\begin{eqnarray}\hspace*{-0.5cm}
H=\sum_j
\Big[-\frac{t}{2}(g_{j+1}^\dagger g^{\;}_j+{\rm H.c.})
-\mu \, n_j+ V \, n_jn_{j+1}\Big].
\end{eqnarray}
The chemical potential $\mu$ fixes $\bar{n}$, the interaction $V>0$ 
is repulsive, and the total number of particles  \(\hat N=\sum_jn_j\) is 
conserved. This bosonic $t$-$V$ model displays a Mott insulating phase. 

As a result of Eq.\,\eqref{btos}, \(H\) has an 
interesting interpretation as an XXZ spin $S=1$ model in a magnetic field,
\begin{eqnarray}
H=\sum_j\Big[-\frac{\bar{n}\, t}{2}(S_j^xS_{j+1}^x +S_j^z S_{j+1}^z) \nonumber\\
-(\mu-2\bar{n}V) S_j^y+ V S_j^y S_{j+1}^y\Big]
\end{eqnarray}
(up to an additive constant). In the following the chemical 
potential will be kept fixed at \(\mu=2\bar{n}V\) in order 
to eliminate the magnetic field term. 

The quantum phase diagram of the $S=1$ XXZ is well known.\cite{ChenXXZ}
For $t=0$, the ground state is antiferromagnetically (Ising) 
ordered in the spin language and in a Mott insulating state in 
the boson language. There is a mass gap in the quasiparticle 
spectrum. According to the Haldane gap conjecture, the mass 
gap remains as the Heisenberg antiferromagnetic line $t=-2V/\bar{n}$ 
is reached. For the bosonic $t$-$V$ model this means that the 
atomic system remains in a Mott insulating state. Next we will 
show that the bosonic $t$-$V$ model is dual to a particle 
non-conserving Hamiltonian with an unconventional superfluid 
ground state.
  
To establish the equivalence of the bosonic $t$-$V$ 
model to a superfluid, we will exploit a duality transformation 
first investigated in the context of spin \(S=1\) models, see 
Ref. \onlinecite{kennedy1991} and references therein. A compact 
expression for the unitary transformation associated to this duality 
is\cite{oshikawa92}
\begin{eqnarray}\label{oshikawa}
\mathcal{U}_{\sf d}=\prod_{j<k}e^{\im \pi S^z_jS^x_k} .
\end{eqnarray}
It follows that 
\begin{eqnarray}
\mathcal{U}_{\sf d}S^x_jS^x_{j+1}\mathcal{U}_{\sf d}^\dagger&=&-S^x_jS^x_{j+1}, \nonumber \\
\mathcal{U}_{\sf d}S^y_jS^y_{j+1}\mathcal{U}_{\sf d}^\dagger&=&S^y_j
e^{\im \pi(S^z_j+S^x_{j+1})}S^y_{j+1},\label{d2} \nonumber \\
\mathcal{U}_{\sf d}S^z_jS^z_{j+1}\mathcal{U}_{\sf d}^\dagger&=&-S^z_jS^z_{j+1}.
\end{eqnarray} 
Therefore, the dual spin model \(H^D=\mathcal{U}^{\;}_{\sf d}\, 
H\, \mathcal{U}_{\sf d}^\dagger\) is
\begin{eqnarray}
H^D=\sum_j\Big[\frac{\bar{n}t}{2}(S_j^xS_{j+1}^x +S_j^z S_{j+1}^z) \nonumber \\
+ V \, S^y_je^{\im \pi(S^z_j+S^x_{j+1})}S^y_{j+1}\Big] .
\end{eqnarray}

The dual model of bosonic atoms is obtained by rewriting the spin 
Hamiltonian \(H^D\) in terms of particle operators. This task is 
easy once the identity 
 \begin{eqnarray}
&&\hspace*{-0.7cm} S^y_je^{\im \pi(S^z_j+S^x_{j+1})}S^y_{j+1}= \nonumber \\
&&\hspace*{-0.2cm} -\frac{1}{4}S_j^y((S_j^+)^2+(S_j^-)^2)((S_{j+1}^+)^2+
(S_{j+1}^-)^2)S_{j+1}^y
\label{useful}
\end{eqnarray}
is established. Then it follows that the dual bosonic $t$-$V$ 
model is described by the Hamiltonian
\begin{eqnarray} \hspace*{-1cm}
H^D&=&\sum_{j}\Big[\frac{\bar{n}t}{2}(g_j^\dagger g^{\;}_{j+1}+g_{j+1}^\dagger g^{\;}_j)
\nonumber \\
&&-\frac{V}{\bar{n}^2}\bar{n}_j(g_j^2+g_j^{\dagger\, 2})(g_{j+1}^2+g_{j+1}^{\dagger\, 2})
\bar{n}_{j+1}\Big], 
\end{eqnarray}
with \(\bar{n}_j=n_j-\bar{n}\). 

The terms 
\begin{eqnarray}
-\frac{V}{\bar{n}^2}\sum_j \bar{n}_j(g_j^2g_{j+1}^{\; \, 2}+
g_j^{\dagger\, 2}g_{j+1}^{\dagger\, 2})\bar{n}_{j+1}. 
\end{eqnarray}
break particle conservation. Bosons are created and annihilated in 
quartets. Hence the U(1) symmetry of particle conservation of the 
$t$-$V$ model is explicitly broken down to 
a discrete \(\mathds{Z}_4\) symmetry  in the dual Hamiltonian. The operator
$\im^{\hat N}=e^{\im \frac{\pi}{2}\sum_jn_j }$
commutes with \( H^D\), and has the interpretation of  counting 
the total number of bosons modulo four.

\end{document}